\documentclass[12pt, a4paper,fleqn,usenatbib]{mnras}

\usepackage{newtxtext,newtxmath}

\usepackage[T1]{fontenc}
\usepackage{ae,aecompl}
\usepackage{ucs}
\usepackage[utf8x]{inputenc}
\usepackage{gensymb}

\usepackage{graphicx}	
\usepackage{amsmath}	
\usepackage{amssymb}
\usepackage{hyperref}
\usepackage{float}

\hypersetup{
	colorlinks=true
}

\title[Vertical metallicity gradients in the Milky Way]{The vertical metallicity 
gradients of mono-age stellar populations in the Milky Way with the RAVE and \textit{Gaia} data}

\author[I. Ciuc\u{a} et al.]
{\parbox{\textwidth}{Ioana Ciuc\u{a}$^{1}$\thanks{E-mail: ioana.ciuca.16@ucl.ac.uk}, Daisuke Kawata,$^{1}$
		Jane Lin$^{2}$,
		Luca Casagrande$^{2}$,
        George Seabroke$^{1}$,
    Mark Cropper$^{1}$}\vspace{0.5cm}
	\\
	$^{1}$ Mullard Space Science Laboratory, University College London,
	Holmbury St. Mary, Dorking, Surrey, RH5 6NT, UK \\
	$^{2}$ Research School of Astronomy \& Astrophysics, Australian National University, Canberra ACT 2611, Australia
}
\date{Accepted 2017 December 12. Received 2017 December 4; in original form 2017 May 30} 

\pagerange{\pageref{firstpage}--\pageref{lastpage}}
\pubyear{2017}

\begin{document}
\label{firstpage}
\pagerange{\pageref{firstpage}--\pageref{lastpage}}
\maketitle

\begin{abstract}
We investigate the vertical metallicity gradients of five mono-age stellar populations between 0 and 11~Gyr for a sample of 18 435 dwarf stars selected from the cross-matched Tycho-Gaia Astrometric Solution (TGAS) and RAdial Velocity Experiment (RAVE) Data Release 5. We find a correlation between the vertical metallicity gradients and age, with no vertical metallicity gradient in the youngest population and an increasingly steeper negative vertical metallicity gradient for the older stellar populations. The metallicity at disc plane remains almost constant between 2 and 8~Gyr, and it becomes significantly lower for the $8 < \tau  \leqslant 11$~Gyr population. The current analysis also reveals that the intrinsic dispersion in metallicity increases steadily with age. We discuss that our results are consistent with a scenario that (thin) disc stars formed from a flaring (thin) star-forming disc.
\end{abstract}




\section{Introduction}
\label{intro}

Galactic archaeology is entering a golden era with the $Gaia$ space mission, which is providing positional and proper motion information for over a billion stars \citep{gaia2016b}. $Gaia$’s Radial Velocity Spectrometer \citep{Cro2011} will provide radial velocity measurements for the 150 million brightest stars in the Milky Way. Large-scale stellar spectroscopic surveys, such as the Radial Velocity Experiment \citep[RAVE;][]{Ste2006}, the Sloan Extension for Galactic Understanding and Exploration \citep[SEGUE;][]{Yanny2009}, the Large Sky Area Multi-Object Fibre Spectroscopic Telescope \citep[LAMOST;][]{Liu2014}, the Apache Point Observatory Galactic Evolution Experiment \citep[SDSS-III/APOGEE;][]{Eis2011, Maj2015}, Gaia-ESO \citep{Gil2012}, the GALactic Archaeology with HERMES \citep[GALAH;][]{Mar2017}, the 4-metre Multi-Object Spectroscopic Telescope \citep[4MOST;][]{deJong2012} and the WHT Enhanced Area Velocity Explorer \citep[WEAVE;][]{Dal2012}, aim to complement the $Gaia$ data with chemical and radial velocity information for the faint stars, thus allowing us to probe the Milky Way to unprecedented scale and detail. The ultimate goal of Galactic archaeology in the Milky Way is to reconstruct the Galaxy's formation and evolution history. For this purpose, one of the most cogent approaches that employ $Gaia$ (and spectroscopic survey data) is to map the chemo-dynamical structure of the Galaxy in chronological order. In this way, we can discern the role of various internal processes in shaping the geometry, dynamics and chemistry of the Galaxy at every evolutionary stage by comparing observations with theoretical models, often in which the time evolution is the clearest indicator.  

Stellar age is an essential ingredient in mapping the chrono-chemo-dynamical structure of the Milky Way. Determining the age of stars is a challenging problem. There are several ways to obtain the ages of stars, including isochrone fitting \citep[e.g.][]{Hol2009, Cas2011, Hay2013}, asteroseismology \citep[e.g.][for a review]{Cha2013} and the C/N abundance ratio \citep{Mas2015, Mar2016}. For example, by using the age information in conjunction with metallicity and distance measurements, we can characterise the radial and vertical metallicity profile of the Galactic disc at every evolutionary stage. Then, we can make informed inferences about the kinematic and dynamical properties of the Galactic disc at the formation and subsequent evolutionary stages.

Using 418 red-giant stars from the CoRoT and APOGEE combined data close to the Galactic plane (6 < $R \lesssim$ 13~kpc, $\left|Z\right|$ < 0.3~kpc, here $R$ and $\left|Z\right|$ are the Galactocentric radius and vertical height from the Galactic plane, respectively), \cite{Cor2017} investigated the radial metallicity distribution of stars in different mono-age stellar populations, defined as a group of stars with similar ages. They found that, at each radius, the dispersion of the metallicity distribution increases with age. Also, their study revealed that the radial metallicity gradient flattens with age. 

These observational trends can be explained as follows. In this paper, we assume that there was no metallicity dispersion for stars formed at a fixed radius as is seen for young B stars in the Solar neighbourhood  \citep{Nie2012, Fel2013}, but there was a negative radial metallicity gradient. Then, at a fixed radius, radial mixing of stars could have brought more metal-rich stars from the inner region and more metal-poor stars from the outer region \citep[e.g.][]{Sel2000, S2009b, S2009, Min2013, Gra2015}, and this would increase the metallicity dispersion. Since older stars had more time to be affected by radial mixing, the older stellar population at a fixed radius contains both more metal-poor and rich stars than the younger population, and this explains the broader metallicity dispersion of the older stars. The flattening of the radial metallicity gradient can be explained by the radial metallicity gradient at formation time being flatter at earlier epochs or radial mixing flattening the radial metallicity gradient for the older stars \citep[e.g.][]{Ros2008, San2009, Min2014, Gra2015, Kaw2017}.

Another study that measured the metallicity distribution in the Galactic disc as a function of age was done by \cite{Xia2015}. They employed the LAMOST Spectroscopic Survey of the Galactic Anti-Center dataset \citep[LSS-GAC;][]{Liu2014}, and used photometric distances to perform isochrone fitting of turn-off stars to obtain stellar ages for 297 042 stars covering a vast region between 7.5 < $R$ < 13.5~kpc and $\left|Z\right|$ < 2.5~kpc. They found that the measured radial metallicity gradients are approximately constant up to the age of 11~Gyr. They considered an age-defined thick disc as the population of stars with age > 11~Gyr and found that this thick disc population has a flat radial metallicity gradient, distinct from the younger thin disc populations.

\cite{Xia2015} also investigated the vertical metallicity gradients of mono-age stellar populations by splitting their sample into age bins of width 2 to 3~Gyr between 2 and 11~Gyr and a wider single age bin for the 11-16~Gyr population. Their results revealed that at the Solar radius, the older mono-age stellar populations display increasingly steep negative vertical metallicity gradients from d[M/H]/d$\left|Z\right| \simeq - \ 0.13$~dex~kpc$^{-1}$  for stars with ages between 2 and 4~Gyr to d[M/H]/d$\left|Z\right| \simeq - \ 0.21$~dex~kpc$^{-1}$  for the 8 - 11~Gyr population. However, the oldest population with age $> 11$~Gyr shows a flatter vertical metallicity gradient of d[M/H]/d$\left|Z\right| = -\ 0.11$~dex~kpc$^{-1}$. Here, we use [M/H] to indicate the metallicity defined as [M/H] = $\rm log_{10}(M/H) - log_{10}(M/H)_{\sun}$, with M indicating the metal mass fraction and H the Hydrogen mass fraction in the star. 

An indirect approach to determine the vertical metallicity of mono-age stellar populations relies on using [$\alpha$/Fe] as a proxy for age \citep[e.g.][]{Sme1992, Hay2013, Hay2015}. Older stars are thought to be more $\alpha$-enhanced than their younger counterparts. For example, using dwarf stars selected from the SEGUE survey, \cite{Sch2014} found a negligible vertical metallicity gradient for different [$\alpha$/Fe] populations above the Milky Way plane (0.3 $< \left|Z\right|< $ 0.6~kpc), where $\alpha$ and Fe indicate $\alpha$-element and iron abundances, respectively. Assuming [$\alpha$/Fe] is a proxy for age, \cite{Sch2014}'s results imply that mono-age stellar populations should have similarly negligible vertical metallicity gradients, in contradiction with the \cite{Xia2015} result. However, a recent numerical simulation by \cite{Min2017} argues that the [$\alpha$/Fe] may not be an irrevocable proxy for the age. What is more, \cite{Mac2017} showed that mono-age stellar populations flare even for the highest [$\alpha$/Fe] \citep[see also][]{Mar2014}. Thus, robust age estimates are needed to map the chemical profile of the mono-age stellar populations of the Galactic disc.

The vertical metallicity gradients as a function of stellar age provide strong constraints on the importance of radial mixing and vertical structure of the star-forming disc at the different epochs. It is a common assumption that the (thin) disc stars formed from a thin high-density molecular gas, where vertical chemical diffusion must be efficient, and thus, the mono-age stellar populations that formed from this gas disc should have no initial vertical metallicity gradients. At the same time, the mono-age stellar populations must have had a negative radial metallicity gradient initially, as was summarised in \cite{Cor2017}, as discussed above. Then, the existence of a vertical metallicity gradient in a mono-age stellar population of the current Galactic disc stars indicates that a dynamical mechanism, such as radial mixing, built up the vertical metallicity gradient \citep[e.g.][]{Min2013, Kaw2017, Sch2017}. Measuring the change of the vertical metallicity gradient with age gives strong constraints on the mechanism. 

Using an N-body simulation of a simplified Milky Way-like disc model and the working assumptions of no initial vertical metallicity gradient and a negative radial metallicity gradient, \cite{Kaw2017} showed that if the mono-age disc population formed with a constant vertical scale height, then radial mixing could drive a positive vertical metallicity gradient of the mono-age stellar population. This process can happen because the metal-rich (poor) stars from the inner (outer) disc tend to end up at a higher (lower) vertical height due to their original higher (lower) vertical energy in the inner (outer) region of the disc. Then, if radial mixing is slow enough, older populations should show a more positive vertical metallicity gradient.  

\cite{Kaw2017} also suggested that a mono-age stellar population can build up a negative vertical metallicity gradient if the scale height of the mono-age disc increases with the radius, i.e. flaring. This is because when the metal-poor stars in the outer region, where stars have higher vertical energy due to their greater scale height, migrate inwards, they oscillate vertically more and become dominant at the higher vertical height in the inner region, which leads to a negative vertical metallicity gradient. In this flaring star-forming disc case, if the radial mixing process is slow enough, the older populations should have a more negative vertical metallicity gradient, because more stars from the outer region migrate into the inner region and occupy the higher vertical height region. 

Hence, the vertical metallicity gradient for the mono-age stellar population provides critical information for understanding the dynamical evolution in the Galactic disc. \cite{Xia2015} did pioneering work in studying the vertical metallicity gradient of the mono-age stellar population. Further research with independent data is needed to follow up on their work. Recently, \cite{gaia2016b, gaia2016a} released their first $Gaia$ data, including the Tycho-Gaia Astrometric Solution dataset \citep[TGAS;][]{Mic2015}, which provides accurate astrometry for more than 2 million bright stars. At the same time, the RAVE survey team made their fifth data release \citep[DR5][]{Kun2017}, consisting of carefully calibrated metallicity and stellar parameters of more than 200 000 stars that are also present in the TGAS which provides accurate parallax, i.e. distance, information. The combined dataset of TGAS and RAVE is a unique dataset that allows us to determine the vertical metallicity gradient. Hence, this paper presents an independent, new measurement of the vertical metallicity gradients as a function of age for Solar neighbourhood stars. Our sample goes up only a few 100~pc from the plane, for which accurate parallax measurements are available. The typical errors in the metallicity measurements with RAVE are about 0.1~dex. This makes it challenging to reliably determine the vertical metallicity gradient of the level of about d[M/H]/d$\left|Z\right|$ $\sim0.1$~dex~$\rm kpc ^{-1}$. Therefore, we employ the Bayesian hierarchical model of \cite{Kel2007} to perform robust linear regression, which is a very flexible method that enables us to measure not only the vertical metallicity gradient and intercept, but also the intrinsic dispersion of the metallicity of mono-age stellar populations. This more sophisticated approach applied to new data allows us to determine for the first time the vertical metallicity gradient of the youngest population of 0-2~Gyr.

The current paper is structured as follows. In Section \ref{method} we describe the observational data, the method employed to obtain the measurements of vertical metallicity gradients for the mono-age stellar population and how the observational bias due to stellar population is accounted for in the analysis. Then, we define mono-age stellar populations as samples of stars belonging to different age bins. We finally describe how we fit the vertical metallicity gradients using the Bayesian hierarchical model of \cite{Kel2007}. Section \ref{results} presents the results of our measurement of the vertical metallicity gradient for the stars with different ages. Section \ref{discussion} provides the discussion and summary of our study.
\section{Methods}
\label{method}

We used the cross-matched TGAS and RAVE DR5 catalogue to use RAVE metal abundance and stellar parameters with the TGAS astrometric data. TGAS is the primary astrometric data set in the first $Gaia$ data release \citep{gaia2016b} and provides position, parallax and proper motion information for 2 057 050 stars which are in the Tycho-2 catalogue \citep{Hog2000}. The parallaxes have median uncertainties of 0.3~mas, with an additional systematic error of around 0.3~mas \citep{gaia2016b, Lin2016}. RAVE is a magnitude-limited (9 < I < 12) spectroscopic survey of stars in the Milky Way using the 1.2 m UK Schmidt Telescope of the Australian Astronomical Observatory (AAO). RAVE DR5 provides effective temperature, surface gravity, and overall metallicity carefully calibrated against other spectroscopic data and the Kepler-2 (K2) asteroseismic data \citep{Kun2017}. From the cross-matched TGAS-RAVE DR5, we obtain parallax and metallicity information for 215 590 stars.

For this study, we select stars with Galactic latitude $\left|b\right| > 25 \degree$ to ensure an unbiased sky coverage by RAVE \citep{Woj2016}. As recommended in \cite{Kun2017}, we select ``normal'' stars with RAVE flags that satisfy c1=c2=c3=c4=...=c20=``n", and with ALGO\_CONV = 0, frac\_c > 0.7, CHISQ < 2000 and SNR > 20 to obtain a sample with high-quality metallicity measurements. These high-quality cuts for metallicity and stellar parameters are complemented by a further cut in relative parallax error less than 20\% for accurate distance measurements in TGAS which allows us to treat the distance errors as Gaussian. As shown below, we will minimise the selection bias induced by the selection by limiting our colour-magnitude range and taking into account stellar population bias. 

Tycho-2 is 90\% complete up to magnitude $V$ = 11.5~mag \citep{Hog2000}. Although the TGAS sample is not the same as the Tycho-2 sample \citep{Are2017}, we assume that TGAS provides a good enough level of completeness for this study in this magnitude limit. Because the TGAS selection against the Tycho-2 should not introduce any bias for the metallicity of stars at different heights which are the focus of this paper, we assume that this is a safe assumption. 

The magnitudes used in this analysis are extinction-corrected using the same method described in Section 10 of \cite{Kun2017}. To control our selection bias, we select samples in $J-K$, $K$ colour-magnitude diagram (CMD).  To obtain accurate age measurements, we focus on dwarf stars in this study. We found that dwarf stars roughly follow $V = K + 3.28(J-K) + 0.35$~mag. This scaling relation provides that $V$ < 11.5~mag corresponds to  $K < 11.15 - 3.28(J-K)$, and this magnitude limit is much brighter than $I$ < 12~mag limit where the RAVE sample is chemically unbiased \citep{Woj2016}. In addition, to sample dwarf stars, we select stars with $0.2 < J - K < 0.4$~mag. As discussed below, this colour cut provides a compromised solution to minimise selection bias and maximise the sample size. 

\begin{figure}
	\centering
	\includegraphics[width=\columnwidth]{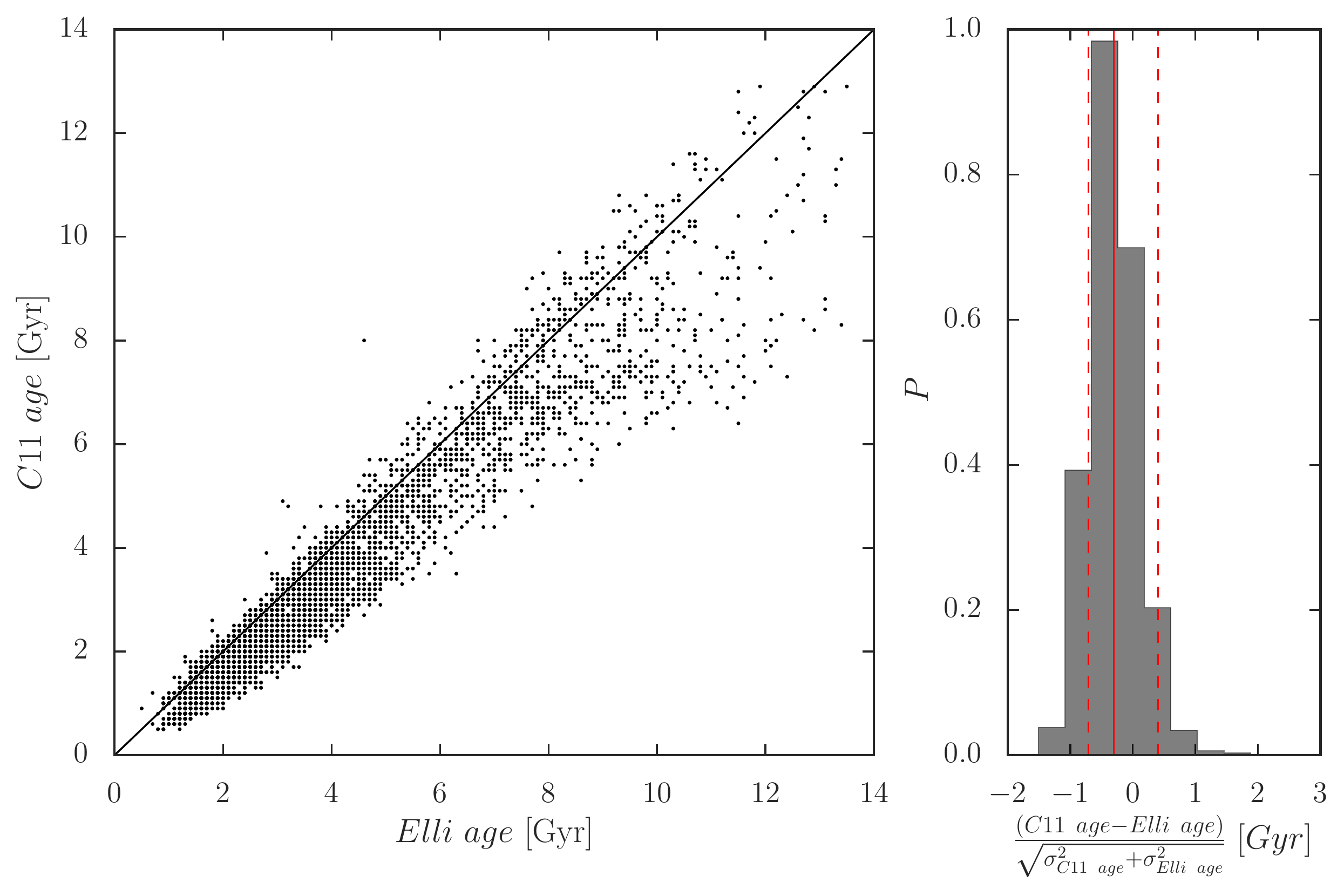}
	\caption{Comparison between stellar ages determinated using $Elli$ and ages in Casagrande et al. (2011)(see text). The scatter in the two estimates is shown in the left panel, with the black line displaying the 1:1 relationship. The distribution of the difference between the two age estimates is shown in the right panel, with the vertical solid and dashed lines indicating the mean of -0.30~dex and standard deviation of 0.40~dex, respectively.} 
	\label{final_fig}
\end{figure}

The stellar age information has been determined using a Bayesian isochrone fitting method called $\tt Elli$ \citep{Linetal17}. The uncertainties of the age measured with $\tt Elli$ are smaller than those reported in the literature due to the extra constraint from TGAS parallaxes \citep{Linetal17}.To assure reliable age estimates, we use only stars whose relative age error is less than 50\%. 

The stellar isochrones used in this version of $Elli$ are taken from the Dartmouth Stellar evolution database \citep{Dot2008}, with $\rm -2.48 \leq [Fe/H]_{init} \leq +0.56$~dex and $\rm -0.2 \leq [\alpha/Fe] \leq +0.8$~dex. The age and mass grids range from 250~Myr to 15~Gyr, and 0.1 to 4.0 $M_{\rm \odot}$, respectively, with solar abundances adopted from \cite{Gre1998}. The equation of state for stellar evolution tracks is assumed to follow the general ideal gas law for stars heavier than 0.8 $M_{\rm \odot}$, with the Debye-H\"{u}ckel correction, and the FreeEOS equation of state for stars less massive than 0.8 $M_{\rm \odot}$ \footnote{http://freeeos.sourceforge.net/}. The conditions at the surface of the star are computed using the \textsc{Phoenix} model atmospheres with $T_{\rm eff}$ ranging from 2000 to 10 000~K and log g from 0.5 to 5.5~dex. For stars hotter than 10 000~K, \cite{Cas2003} model atmospheres are employed.

The method consists of sampling from a posterior distribution $p_{\rm 1}$ which is the product of a likelihood $L$ and the prior distribution $p_{\rm 0}$ to construct a probability distribution for the parameters of interest, which in this case, are the age ($\tau$), mass ($M$), and bulk metallicity $\rm [Fe/H]_{\rm init}$. The mathematical description for the posterior probability distribution is given as:
\begin{eqnarray}
\label{eq1}
p_1(\tau, M, \mbox{\rm [Fe/H]}_{\rm init} | T_{\rm eff}, \mbox{\rm [Fe/H]}, d) \propto p_{\rm 0} L,
\end{eqnarray} where the likelihood $L$ is defined as:
\begin{eqnarray}
\label{eq2}
L = \prod_{i}\frac{1}{\sqrt{2\pi}\sigma_{\rm i}} \times exp\left (-\frac{(O_{\rm i} - S_{\rm i})^2}{2\sigma_{\rm i}^2}\right),
\end{eqnarray}
where $O_{\rm i}$ is the observed values, $T_{\rm eff}$, [Fe/H] and $M_{\rm K}$, and $S_{\rm i}$ are model values. Note that the absolute magnitude, $M_{\rm K}$, depends on the distance, $d$. 
 
Similar to our main analysis of determining the vertical metallicity gradient described below, the posterior is sampled using Markov-Chain Monte Carlo (MCMC). The observed values, $\rm O_{\rm i}$, are compared with model values, $\rm S_{\rm i}$,  obtained by interpolating $\tau, M, {\rm [Fe/H]_{\rm init}}$ over the corresponding grid values of isochrones. We did not use logg due to its large uncertainty in the RAVE data. $M_{\rm K}$ and the distance were obtained from the 2MASS apparent magnitude and the TGAS parallax, respectively. For each isochrone, the likelihood is then computed following equation (\ref{eq2}), assuming the observed parameters have Gaussian errors and the age estimate is obtained by maximising the logarithm of the likelihood function for a given metallicity.

To validate the version of $Elli$ used in this paper, we used the $Elli$ age measurements for the sample of \cite{Cas2011}, and Fig. \ref{final_fig} provides a comparison between the ages derived with and the ages shown in \cite{Cas2011}. The difference in the ages between two methods with respect to their errors are shown in the right panel and are indicative of a reasonable agreement. Although the $Elli$ age is systematically larger than \cite{Cas2011}, the differences are within error, which is not surprising given that different isochrones are applied in these two methods. In Appendix \ref{appendix}, we also provide a more extensive comparison with another isochrone fitting method \citep{PCM2017}. The results in Appendix \ref{appendix} also demonstrate that the age measurements suffer from systematic uncertainty. However, it proves that the qualitative trends discussed in this paper are robust.

\begin{figure}
	\centering
	\includegraphics[width=\columnwidth]{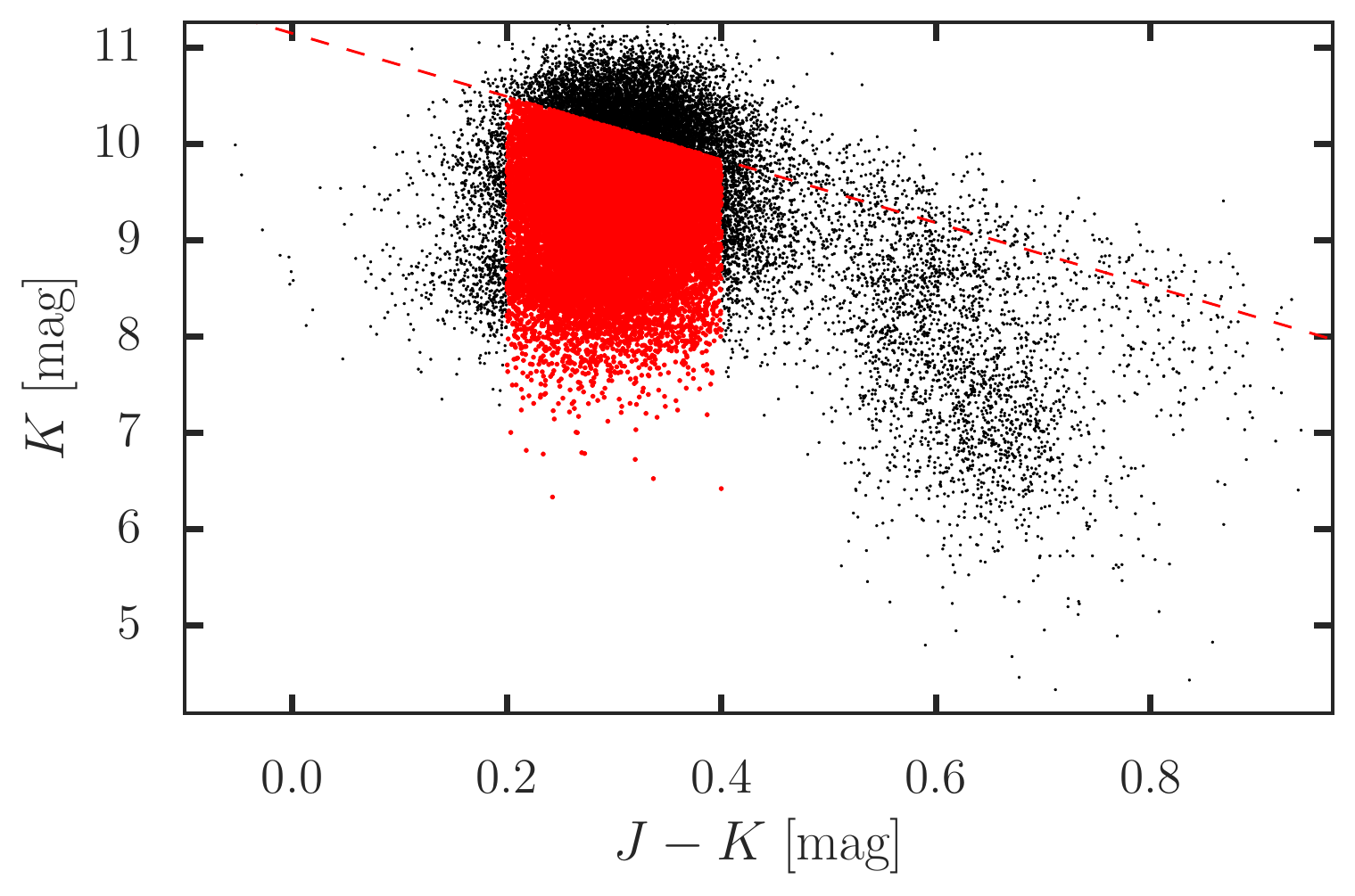}
	\caption{Colour-magnitude diagram of the selected sample before applying a cut in relative age error (red dots) and of the matched TGAS-RAVE DR5 sample (black dots) at $\left|b\right| > 25 \degree$. The magnitude cut $K<11.15 - 3.28(J-K$)~mag is shown as the dotted red line.}
	\label{cmd}
\end{figure}

\begin{figure}
	\centering
	\includegraphics[width=\columnwidth]{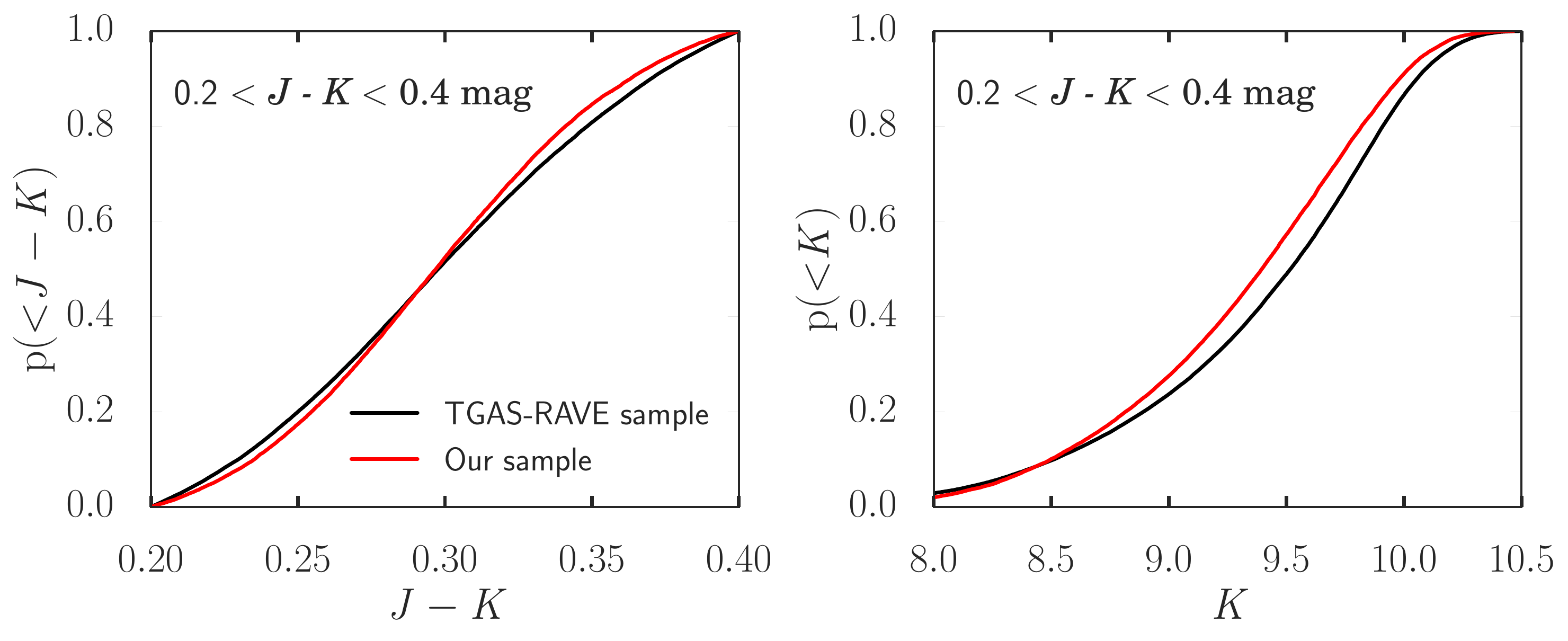}
	\caption{Cumulative distribution functions for the $J-K$ colour (left) and $K$ magnitude (right) between our selected sample before applying a cut in relative age error (red) and all the TGAS and RAVE cross-matched data, i.e. the pre-selection data (black) in the selected magnitude range and with a colour selection of 0.2 < $J-K$ < 0.4~mag. }
	\label{jk}
\end{figure}

\begin{figure}
	\centering
	\includegraphics[width=\columnwidth]{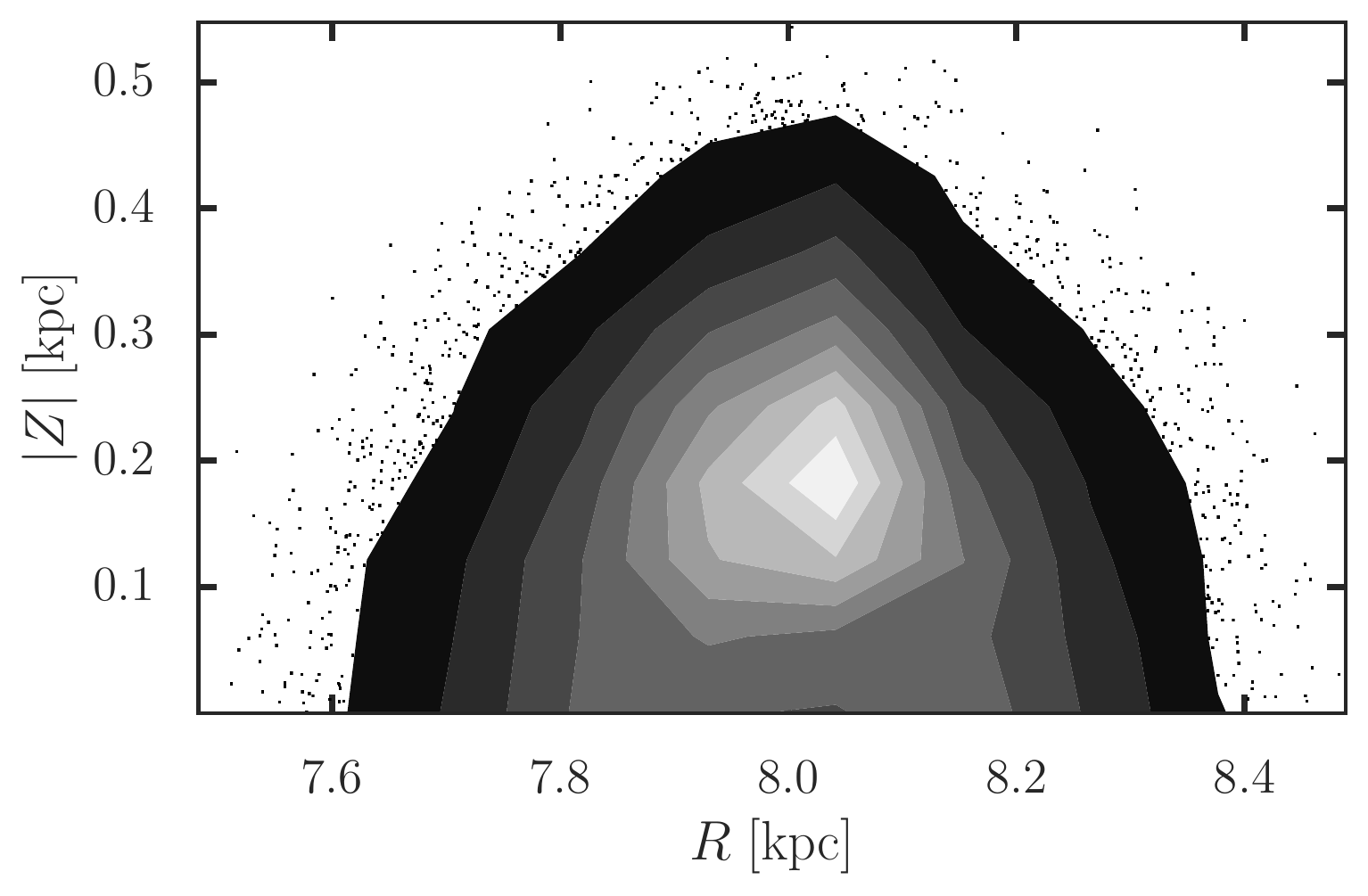}
	\caption{Spatial distribution in the $R-\left|Z\right|$ plane for our selected sample. The greyscale contours indicate the number density distribution of stars, with white and black corresponding to the the highest and lowest density region, respectively. }
	\label{rgal}
\end{figure}

Fig. \ref{cmd} shows the $J-K$ vs.\ $K$ colour-magnitude diagram for our selected sample (red) before applying a cut in relative age error and the matched TGAS-RAVE DR5 (black) in the selected sky coverage of $\left|b\right| > 25 \degree$. The left and right panels of Fig. \ref{jk} display the cumulative distribution functions in the $J-K$ and $K$ distribution, respectively, for our selected sample before applying a cut in relative age error and the TGAS and RAVE cross-matched sample (referred as ``TGAS+RAVE sample") within the chosen colour ($0.2 < J-K < 0.4$~mag) and magnitude range. Fig. \ref{jk} shows that for $0.2 < J-K < 0.4$~mag, the cumulative distribution function of the $J-K$ distribution for our sample follows that of the RAVE-TGAS sample where no selection is applied. The cumulative distribution function of the $K$ distribution for our sample shows no significant deviation that of the RAVE+TGAS sample. Although the $0.2 < J-K < 0.4$~mag seems to introduce bias, this is our compromised solution after we explored various other colour cut ranges. Our final data set after applying a cut in relative age error includes 18 435 sources populating a limited volume in both the vertical and radial direction of $\sim 500$~pc from the Sun as shown in Fig. \ref{rgal}.

\begin{figure*}
	\centering
	\includegraphics[width=\textwidth]{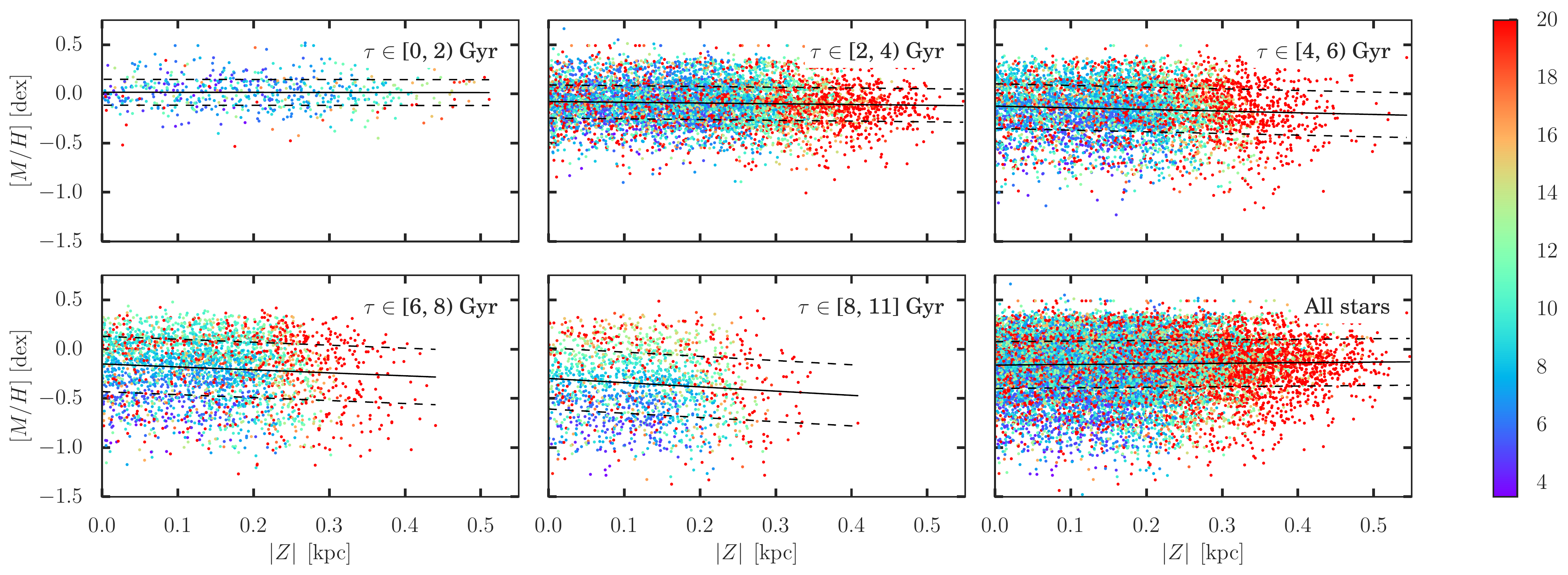}
	\caption{The metallicity distribution with height for the five mono-age stellar populations considered and the sample for the whole age range, including stars older than 11~Gyr (bottom right). Points are colour-coded by the stellar population weight as defined in Sec. \ref{method} and interpreted in Sec. \ref{results}. The lower is the probability to observe a star, the redder is its colour, and the higher is its weight. For each subplot, overplotted in solid black line is the best-fit linear regression slope from the unweighted case. The dotted black lines represent the dispersion.}
	\label{full}
\end{figure*}

We analyse the vertical metallicity gradient for five mono-age stellar populations, defined as populations of stars in 2-3~Gyr age bins. In this paper, we use age intervals of 0-2, 2-4, 4-6, 6-8 and 8-11~Gyr and call them mono-age stellar populations. Strictly speaking, a 2-3~Gyr age bin cannot be called a mono-age stellar population. However, since we need a statistically significant sample of stars for each population, we use this large age bin. The cut in age at 11~Gyr is also due to the same sample size requirement (see Appendix \ref{appendix}).

The vertical metallicity gradients of the mono-age stellar populations are analysed using the linear regression method of \cite{Kel2007}, which employs a Bayesian hierarchical model. This approach enables us to measure the vertical metallicity gradient taking into account heteroscedastic measurement errors and intrinsic scatter in the regression relationship \citep{Kel2007}. The equations to describe the probabilistic hierarchical model are given in \cite{Kel2007} as:
\begin{eqnarray}
\xi_{\rm i} & \sim & p(\xi|\psi) \label{eq-hierxi} \\
\eta_{\rm i}|\xi_{\rm i} & \sim & N(\alpha + \beta \xi_{\rm i}, \sigma^2) \label{eq-hiereta} \\
y_{\rm i},x_{\rm i}|\eta_{\rm i},\xi_{\rm i} & \sim & N_2([\eta_{\rm i},\xi_{\rm i}],\Sigma_{\rm i}) \label{eq-hierxy}
\end{eqnarray}
where $(\xi_{\rm i}, \eta_{\rm i})$ are the actual values. The observed values $(x_{\rm i}, y_{\rm i})$ are measured with errors ($\epsilon_{\rm x, i}, \epsilon_{\rm y, i}$) 
as:
\begin{eqnarray}
x_{\rm i} & = & \xi_{\rm i} + \epsilon_{\rm x,i} \label{eq-xerr} \\
y_{\rm i} & = & \eta_{\rm i} + \epsilon_{\rm y,i} \label{eq-yerr}.
\end{eqnarray}

In this hierarchical model, $\xi_{\rm i}$ is drawn from a probability distribution described by a set of parameters $\psi$. $\eta_{\rm i}$ is drawn from a normal distribution with mean of $\alpha + \beta \xi_{\rm i}$ and variance $\sigma^2$, where $\alpha$, $\beta$,  $\sigma$ are the slope, intercept and intrinsic dispersion, respectively. The observed values, $x_{\rm i}$ and $y_{\rm i}$ come from the multivariate normal density with mean $\xi_{\rm i}$ and $\eta_{\rm i}$ and covariance matrix $\Sigma_{\rm i}$, where $\Sigma_{\rm 11, i} = \sigma_{\rm y,i}^2$, $\Sigma_{\rm 22, i} = \sigma_{\rm x,i}^2$, where $\sigma_{\rm x,i}, \sigma_{\rm y,i}$ are the measurement errors on $x$ and $y$ for the data point $\rm i$, respectively. The non-diagonal terms in $\Sigma_{\rm i}$ are zero since we assume non-correlated errors between metallicity and distance for our data. The method is based on deriving a likelihood function for the measured data $p(x_{\rm i},y_{\rm i}|\theta,\psi) = p(y_{\rm i}|x_{\rm i},\theta,\psi) p(x_{\rm i}|\psi)$, where $\rm \theta = (\alpha, \beta, \sigma^2)$ \citep{Kel2007}. In the current analysis, we model $\xi$ as a Gaussian distribution, characterised by a mean $\mu_{\rm 1}$ and variance $\tau_{\rm 1}$. These parameters define the parameter set $\psi$ \citep{Kel2007}.

For each mono-age stellar population sample, we obtain the vertical height of stars $\left|Z\right|$ from a distance calculated from the TGAS parallaxes and Galactic latitudes, assuming the vertical offset of the Sun to be $Z_{\sun} = 25$~pc \citep{Bla2016}. The vertical distance errors are computed as $sin(b) \pi_{\rm e}/\pi^{2}$, where $\pi$ is the parallax and $\pi_{\rm e}$ is the error in parallax from the TGAS data, with a 0.3~mas systematic added in parallax \citep{gaia2016a}. The distance errors are not simply Gaussian with dispersion of $\pi_{\rm e}/\pi^2$ \citep[e.g.][]{Bai2015}. However, for simplicity we assume Gaussian errors to be a good approximation for our selected data with $\pi_{\rm e}/\pi < 0.2$ and $\pi > 0.0$. The metallicity and their associated errors are taken from the RAVE DR5 catalogue.

In the probabilistic model, we also consider the effect of the stellar population bias as suggested in \cite{Cas2016}. These probabilities are determined as a function of stellar age ($\tau$), metallicity ([M/H]) and distance ($d$), assuming the Salpeter Initial Mass Function \citep{Sal1955}, using the approach of \cite{Cas2016} shown schematically in their Fig. 11. To avoid an extrapolation which has a risk to cause too extreme values, we use the probability associated with $\rm [M/H] = 0.5$ and $\tau = 13$~Gyr for those stars with $\rm [M/H] > 0.5$ or $\tau > 13$~Gyr, respectively. We then define a weight as $w_{\rm i} = {p_{\rm i}({\tau_{\rm i}, {\rm [M/H]_i}, d_{\rm i}})}^{-1}$ for each star, which is then used to weight the log of the likelihood functions: $\frac{Nw_i{\rm i}}{\sum_{i}w_{\rm i}}\ln L_{\rm i}({\rm [M/H]}| |Z|, \theta, \psi)$ and $\frac{Nw_{\rm i}}{\sum_{i}w_{\rm i}} \ln L_{\rm i}(|Z| | \psi)$, where $\ln L_{\rm i}({\rm [M/H]})$ and $\ln L_{\rm i}(|Z|)$ correspond to $p(y_{\rm i}|x_{\rm i},\theta,\psi)$ and $p(x_{\rm i}|\psi)$, respectively. $N$ is the number of stars in the sample. This entropic weighting, i.e. weighting in the log for both likelihood functions is allowed since  $\rm [M/H]$ and $\left|Z\right|$ are assumed to be independent measurements. For simplicity, we do not take into account the uncertainties associated with the weights, and instead use the weights from the measured age, [M/H] and distance, i.e. $1/\pi$, without considering their errors. To avoid an unreasonably large weight value due to a very low probability, which can happen if the measured values are significantly different from their actual value, we cap the weight at $w_{\rm i} < 20.0$, corresponding to a probability $p_{\rm i} > 0.05$. In this paper, we will refer to this weight as the stellar population weight and will perform an analysis with and without the weights. It is important to note that the stellar population weight does not correct the stellar population bias perfectly. The computed probabilities are dependent on the isochrones adopted in the stellar population model, and the current analysis does not take into account the observational errors of the assigned probabilities. Therefore, the results of the weighted case are intended to provide the effect of the stellar population bias in a qualitative manner. 

Inference of the parameters is explored using a MCMC approach with an Adaptive Metropolis-Hastings algorithm, which allows us to sample posterior distributions and provides the full marginalised probability of the parameters. We implement this method in {\tt pymc}, which facilitates modelling the complex dependencies of the priors employed in the current analysis. The prior space consists of the model parameters, $\alpha, \beta, \sigma$, as well as those describing the $\xi$ distribution, and a full account is given in \cite{Kel2007}. They are drawn from the distributions considered in \cite{Kel2007} apart from $\mu$ and $\tau$, which are sampled from uniform distributions with more restrictive bounds.
\section{Results}
\label{results}
\begin{table*}
	\caption{Results from the linear regression model.}
	\label{table}
	\begin{tabular}{lccccccc}
		\hline
		&  \multicolumn{3}{c}{unweighted} & & \multicolumn{3}{c}{weighted} \\
		\cline{2-4} \cline{6-8} 
		Age &  d[M/H]/d$\left|Z\right|$ & $\rm \beta$ & $\rm \sigma$ & &
		d[M/H]/d$\left|Z\right|$ & $\rm \beta$ & $\rm \sigma$  \\
		Gyr  & dex kpc$^{-1}$ & dex  & dex & & 
		dex kpc$^{-1}$ & dex & dex \\
		\hline

		0 - 2 &  $-0.007^{0.070}_{0.070}$  & $0.016^{0.015}_{0.015}$ & $ 0.131^{0.006}_ {0.006}$ &&
		$0.014^{0.070}_{0.068}$ & $ 0.003^{0.016}_{0.017}$ & $0.144^{0.006}_{0.006}$  \\
		\\ 
		2 - 4 &  $-0.076^{0.024}_{0.023}$ & $-0.078^ {0.005}_ {0.005}$ & $0.168^{0.002}_ {0.002}$ &&
		$-0.095^{0.021}_{0.023}$ & $-0.075^{0.005}_{0.005}$ & $0.174^{0.002}_{0.002}$  \\
		\\
		
		4 - 6 &  $-0.169 ^{0.037}_ {0.037} $  & $-0.125^ {0.007}_ {0.006}$ & $ 0.227^ {0.003}_ {0.002}$ &&
		$-0.192^{0.036}_{0.035}$ & $-0.111^{0.007}_{0.007} $ & $0.232^{0.003}_{0.003}$  \\
		\\
		6 - 8 &  $-0.302^{0.076}_{0.075}$ & $-0.150^{0.012}_{0.012}$ & $0.281^{0.004}_{0.004}$ &&
		$-0.411^{0.068}_ {0.070}$ & $-0.113^{0.012}_ {0.012}$ & $0.287^{0.004}_{0.004}$ \\
		\\
		
		8 - 11 &  $-0.429^ {0.120}_ {0.127}$  & $-0.298^ {0.017}_ {0.018}$ & $ 0.311^{0.006}_{0.006}$ &&
		$-0.653^{0.117}_{0.113}$ & $-0.218^{0.017}_{0.018}$ & $0.321^{0.007}_{0.007}$  \\ 
		\\
		All stars &  $0.060^ {0.021}_{0.022}$  & $-0.162^{0.004}_{0.004}$ & $0.238^{0.002}_{0.001}$ &&
		$-0.026^{0.020}_{0.021}$ & $-0.140^{ 0.004}_{0.004}$ & $0.239^{0.002}_{0.002}$  \\
		\hline
	\end{tabular}
\end{table*}

In Fig. \ref{full}, we present the vertical distribution in metallicity of the mono-age stellar populations and of the full sample in the whole age range of $0<{\rm age}\leq 11$~Gyr. As discussed in Section \ref{method}, in our selected colour-magnitude range each star has a different probability of being observed in the chosen colour and magnitude range depending on distance, age, and metallicity. We compute the stellar population weight as the inverse of this probability and colour-code the points in Fig. \ref{full} by the stellar population weight. We apply the Bayesian hierarchical regression model described in Section \ref{method} to the data for each mono-age stellar population. This method is employed for two cases, namely with and without the stellar population weight, referred to as the stellar population bias ``weighted" and ``unweighted" case, respectively. Fig. \ref{full} also shows our best fit linear regression slope for the stellar population unweighted case as the median of the posterior distribution of the associated, marginalised MCMC chains. Lower and upper uncertainties correspond to the 16th and 84th quartile, respectively. Table \ref{table} gives the best-fit values with uncertainties for the slope, $\alpha={\rm d[M/H]/d}\left|Z\right|$, intercept, $\beta$, and dispersion, $\sigma$, for the five mono-age populations as well as for the entire sample, including stars with ages greater than 11~Gyr.

\begin{figure}
	\centering
	\includegraphics[width=\columnwidth]{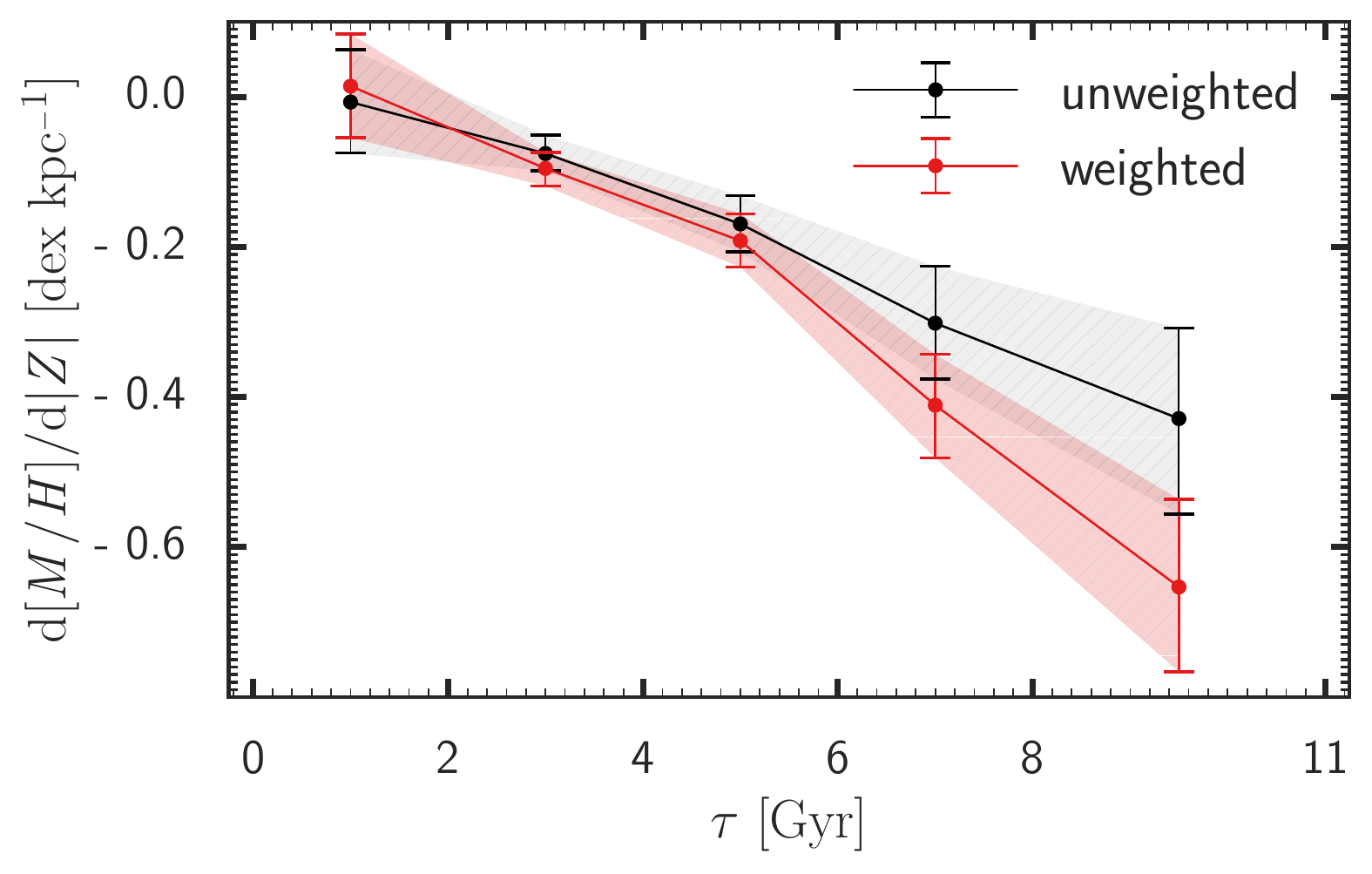}
	\caption{The best-fit estimates and their uncertainties for the slope of vertical metallicity gradient as a function of age, $\tau$,  in both the weighted (light red) and unweighted (black) cases. The shaded area gives the uncertainity space of the slope in the unweighted case.}
	\label{mhz}
\end{figure}

\begin{figure}
	\centering
	\includegraphics[width=\columnwidth]{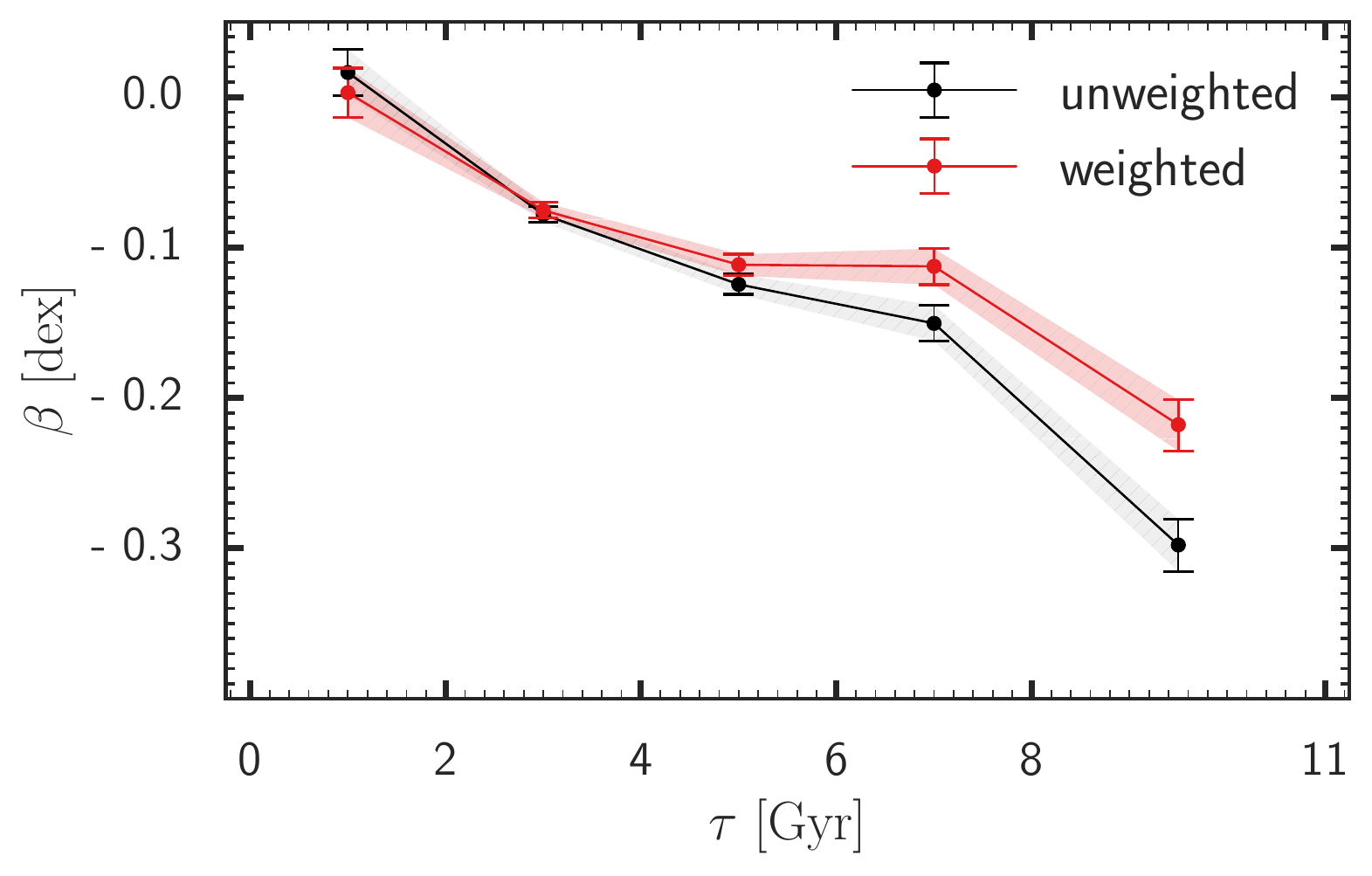}
	\caption{The best-fit estimates and their uncertainties for the intercept, $\beta$, from our vertical metallicity gradient fitting as a function of age, $\tau$,  in both the weighted (light red) and unweighted (black) cases. The shaded area gives the uncertainity space of the intercept in the unweighted case.}
	\label{intercept}
\end{figure}

The slope for each mono-age stellar population as a function of age is shown in Fig. \ref{mhz}. The results of the current analysis revealed that all our mono-age stellar populations show a negative vertical metallicity gradient. There appears to be a steady increase in steepness of the negative slope with increasing age. The youngest population of stars with ages between 0 and 2~Gyr show an almost flat vertical metallicity gradient, d[M/H]/d$\left|Z\right|$ = $-0.013^{0.066}_{0.070}$~dex~kpc$^{-1}$. The oldest population reach a significantly more negative value, at d[M/H]/d$\left|Z\right|$ = $-0.441^ {0.122}_ {0.125}$~dex~kpc$^{-1}$ in the case of the stellar population unweighted case. The stellar population weighted case reveals a qualitatively similar trend of negatively steeper vertical metallicity gradient. The stellar population weighted case gives a slightly steeper slope than the unweighted case, especially for the relatively older populations, $\tau \gtrsim 4$~Gyr. As it can be seen in Fig.~\ref{full}, at lower $\left|Z\right|$, the stellar population weight for the higher metallicity stars is higher than the stellar population weight for the lower metallicity stars. As a result, higher weights for higher metallicity stars lead to a steeper negative slope and higher values for the intercept (Fig. \ref{intercept}). Even with the added stellar population bias effect, the current results reveal an almost flat vertical metallicity gradient for the youngest population and increasingly negative vertical metallicity gradient for the older age populations. Hence, we conclude that the observed trend is robust and likely not due to the observational selection or stellar population biases.

Fig. \ref{intercept} reveals the trend for our best-fit linear regression intercept values, $\beta$, of each mono-age stellar population for both the stellar population weighted and unweighted cases. The stellar population unweighted and weighted cases show a similar trend. The youngest stellar population of $0 < \tau < 2$~Gyr has a systematically higher metallicity than the older populations. The stars with ages between 2 and 8~Gyr show similar values for the intercept. The change in the intercept values is smaller for the stellar population weighted case. On the other hand, the oldest stellar population, $8< \tau<11$~Gyr, shows significantly lower metallicity. The intercept indicates the metallicity at the disc plane, $\left|Z\right| = 0$~kpc. The stellar population weighted case shows systematically higher values of the metallicity at the plane for the stars older than $\sim4$~Gyr. This trend corresponds to the steeper negative slope for the stellar population weighted case in Fig.~\ref{mhz}. The steeper negative slope, driven by the higher weight of higher metallicity stars at lower $\left|Z\right|$, leads to the higher [M/H] at the disc plane for the stellar population weighted case. The similarity between the weighted and unweighted cases reinforces the robustness of our results. In fact, this trend is reminiscent to the age-metallicity relation observed for the Solar neighbourhood stars \citep[e.g.][]{Cas2011, Hay2013, Cor2017}. For example, Fig. 9 of \cite{Hay2013} shows stars younger than 2 Gyr have higher [Fe/H] than older stars \footnote{Note that \cite{Hay2013} used [Fe/H]. On the other hand, we use the [M/H] value from RAVE DR5. Therefore, we use [Fe/H] when quoting the results of \cite{Hay2013}.}. This systematically higher metallicity for stars younger than $\sim 2$~Gyr is also seen in fig. 4 of \cite{Cor2017}. \cite{Hay2013} also showed that stars older than 8~Gyr have lower [Fe/H] systematically, and identified them as the age-defined thick disc.

\begin{figure}
	\includegraphics[width=\columnwidth]{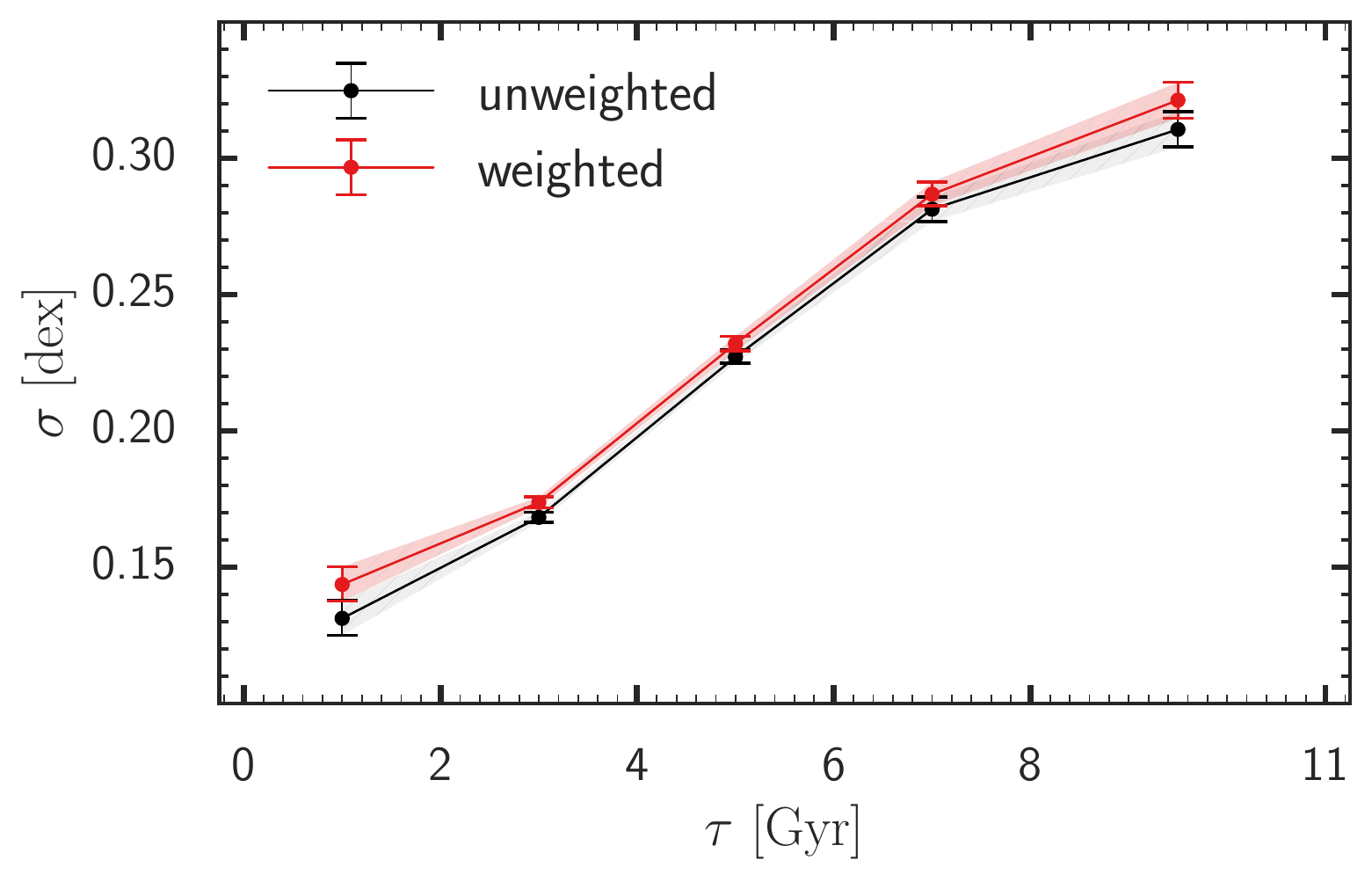}
	\caption{The best-fit estimates and their uncertainties for the dispersion, $\sigma$, for our vertical metallicity gradient fitting for mono-age stellar populations as a function of age, $\tau$, in both the weighted (light red) and unweighted (black) cases. The shaded area gives the uncertainity space of the dispersion in the unweighted case. }
	\label{dispersion}
\end{figure}

Finally, the values of the dispersion, $\sigma$, from our best-fit vertical metallicity gradient model for the mono-age stellar populations for both the stellar population weighted and unweighted cases are shown in Fig. \ref{dispersion}. These values are almost the same for both the weighted and unweighted cases. Our results reveal that the intrinsic scatter increases steadily with age, from $\sigma \sim 0.13$~dex for the youngest population to $\sigma \sim 0.31$~dex for the oldest population. As discussed in Section \ref{intro}, the larger intrinsic dispersion of the metallicity distribution of the older stellar populations can be explained by a radial mixing mechanism which brings more metal-rich (poor) stars from the inner (outer) region of the Galactic disc. The steady increase of the dispersion indicates that radial mixing has not reached saturation, which is expected to happen if the radial mixing processes mix stars born at different radii quickly. It is also important to note that our metallicity dispersion is systematically higher than what is obtained by \cite{Cor2017} for their mono-age stellar populations. This may be due to the large measurement uncertainties in [M/H] from the RAVE data, as compared to their metallicity from high signal-to-noise data from the APOGEE data. However, since there should not be a correlation between the age and metallicity errors, e.g. older stars should not have higher uncertainties in the RAVE data, we conclude the trend that we observe is not induced by the observational uncertainties or selection biases. Instead, it is likely to be due to an inherently higher dispersion in metallicity for the older population. 

\section{Discussion and Summary}
\label{discussion}

In this paper, we measured the dependence of metallicity on vertical height for five different mono-age stellar populations by analysing combined data from the TGAS and RAVE DR5 catalogues with an advanced linear regression model. Our main results are summarised as follows:

\begin{enumerate} 

\item The youngest population has an almost flat vertical metallicity gradient. Contrastingly, the older population shows a negative vertical metallicity gradient, i.e. lower [M/H] at a higher height. Furthermore, the mono-age stellar populations display an increasingly steeper negative vertical metallicity gradient with age. 

\item From the intercept value from the linear regression model, which yields the metallicity at the disc plane (Z= 0~kpc), we found an age-metallicity relation. The youngest mono-age stellar population with ages between 0 and 2~Gyr has systematically higher metallicity than stars older than 2~Gyr. Across the three mono-age stellar populations within the age range of 2 to 8~Gyr, the metallicity at the disc plane remains almost constant. On the other hand, the oldest population of $8 < \tau  \leqslant 11$~Gyr shows significantly lower metallicity. 

\item The linear regression model employed in our study enables us to analyse the intrinsic dispersion in the metallicity distribution. We find that the dispersion increases steadily with age up to 8 - 11~Gyr. 

\end{enumerate}

We minimised the observational selection bias induced by the various selection criteria by selecting a sample in a specific colour and magnitude range. Additionally, we applied the stellar population bias correction by weighting the likelihood function with the inverse of the probability of seeing a star having a particular age and metallicity and located at a certain distance in the selected colour and magnitude range. In a qualitative sense, the results obtained by taking into account the stellar population bias are similar to those obtained when no stellar population weighting was applied. This fact is evidence that the results of this analysis are robust, and unlikely induced by observational selection bias or stellar population bias.

The vertical metallicity gradients found in our study are qualitatively consistent with the study by \cite{Xia2015} despite their vertical range (0 < $\left|Z\right|$ < 2~kpc) being much more extended than ours. At the Solar radius, their results reveal a steeper negative vertical metallicity gradient for the older mono-age stellar populations, which is in qualitative agreement with our results for stars with ages between 2 and 11~Gyr. \cite{Xia2015} also found that stars with ages between 11 and 16~Gyr have a flatter gradient. Unfortunately, we cannot measure the vertical metallicity gradient of this mono-age population since we do not have a statistically significant subset of stars older than 11~Gyr in our sample. The bright magnitude limit and the narrow colour range employed in our selection, as well as the limited volume coverage of our sample, i.e. we are still rather confined to the plane, where the majority of stars are young, limit our sample to stars no older than 11~Gyr. At the youngest end, we found that stars with ages between 0 and 2~Gyr have an almost flat vertical metallicity gradient. Since \cite{Xia2015} did not show the vertical metallicity measurement for stars younger than 2~Gyr, our results reveal for the first time the flat vertical metallicity gradient of the youngest population, and connect it with the gradually steeper negative vertical metallicity gradient of the older population.

Before discussing the various scenarios that can explain our results, it is important to consider the Galactic disc evolution scenarios that our results can reject. The fact that the older population has a distinctly negative vertical metallicity gradient and the gradient becomes steeper with age strongly disfavors the existence of a very strong vertical phase mixing of stars (like strong heating by satellite accretion). Such a process would have driven a flat vertical metallicity gradient across the mono-age stellar populations formed before the heating happened. In the light of our results, the Milky Way disc should not have undergone dramatic heating leading to strong vertical mixing. Furthermore, they seem to support the recent claims that the Milky Way has not undergone any major merger in the last 10~Gyr \citep{Ruc2015, Cas2016}.

As justified in Section \ref{intro}, it is natural to assume that each mono-age disc population had a negative radial metallicity gradient and no vertical metallicity gradient when they formed. Under these conditions, our observed trend challenges one of the numerical results of \cite{Kaw2017}. According to their study, if a mono-age stellar population of the Galactic disc had a constant scale height at all radii, radial mixing would have driven a positive vertical metallicity gradient as discussed in Section \ref{intro}. Our results invalidate constant scale height for the initial structure of the mono-age disc populations and suggest that the Galactic disc stars should not have had a star-forming disc with a constant scale height across all radii.

On the other hand, our results support another model of \cite{Kaw2017} where they consider that each mono-age stellar population formed from a flaring star-forming disc. Flaring star-forming regions have been observed previously in the Milky Way in the distribution of the HI and molecular gas \citep[e.g.][]{Sco1993, Nak2016}, and young stellar populations traced by Cepheids \citep{Fea2014}. It is not impossible to expect that all the thin disc stars formed in a flaring star-forming region \citep[e.g.][]{Rah2014, Min20151}. \cite{Kaw2017} demonstrate that if this flaring thin disc has a negative radial metallicity gradient when the mono-age stellar populations formed, then radial mixing processes could lead to a negative vertical metallicity gradient as discussed in Section \ref{intro}. If radial mixing was a slow process, then this scenario could explain our observational trend.

Based on the same flaring disc scenario, if the star-forming region had a similar level of flaring across all mono-age stellar populations, but a steeper negative d[M/H]/dR at an earlier epoch, radial mixing could drive the steeper vertical metallicity gradient for the older mono-age stellar populations. However, \cite{Cor2017} found an opposite trend in the radial metallicity gradient of the current disc, with stars with ages between 1 and 4~Gyr having d[M/H]/d$R$ $\simeq-$ 0.07~dex~kpc$^{-1}$, whereas the older population shows a flatter gradient of d[M/H]/d$R$ $\simeq-$ 0.03~dex~kpc$^{-1}$. A similar trend was observed by \cite{Cas2011} when they analysed radial metallicity gradients in the Solar neighbourhood using the guiding radius of stars. We still do not know whether the flatter radial metallicity gradient for the old disc is due to the radial metallicity gradient of the star-forming disc being flatter at an earlier epoch, or due to radial mixing flattening the radial metallicity gradient more prevalently for the older disc. Thus, this raises the question about what was the initial radial metallicity gradient of each generation of stars in the Galactic disc. An extensive set of stars with accurate measurements of age and metallicity spanning a large vertical and radial range should provide stronger constraints on the initial metallicity distribution of the Galactic disc at different epochs and the strength of radial mixing effects by comparing with numerical simulations. Such data will be available soon with $Gaia$ DR2 complemented by high-resolution spectroscopic survey data, such as APOGEE-2, $Gaia$-ESO and GALAH, and asteroseismic age information from the K2 campaign \citep[e.g.][]{Ste2015} and NASA's TESS mission, and ultimately by ESA'S Plato mission.
\section*{Acknowledgements}
IC, DK and MC acknowledges the support of the UK's Science \& Technology Facilities Council (STFC Grant ST/K000977/1 and ST/N000811/1). IC is also grateful the STFC Doctoral Training Partnerships Grant (ST/N504488/1). LC is supported by Australian Research Council Future Fellowship FT160100402.

This work has made use of data from the European Space Agency (ESA)
mission {\it Gaia} (\url{https://www.cosmos.esa.int/gaia}), processed by
the {\it Gaia} Data Processing and Analysis Consortium (DPAC,
\url{https://www.cosmos.esa.int/web/gaia/dpac/consortium}). Funding
for the DPAC has been provided by national institutions, in particular
the institutions participating in the {\it Gaia} Multilateral Agreement.

Funding for RAVE has been provided by the Australian Astronomical
Observatory; the Leibniz-Institut fuer Astrophysik Potsdam
(AIP); the Australian National University; the Australian Research
Council; the French National Research Agency; the German
Research Foundation; the European Research Council (ERC-StG
240271 Galactica); the Istituto Nazionale di Astrofisica at Padova;
The Johns Hopkins University; the National Science Foundation
of the USA (AST-0908326); the W. M. Keck foundation; the Macquarie
University; the Netherlands Research School for Astronomy;
the Natural Sciences and Engineering Research Council of Canada;
the Slovenian Research Agency; the Swiss National Science Foundation;
the Science \& Technology Facilities Council of the UK; Opticon;
Strasbourg Observatory and the Universities of Groningen,
Heidelberg and Sydney. The RAVE web site is at http://www.ravesurvey.org

We thank Jo Bovy for making $\tt GALPY$ publicly available \citep{galpy} and to {\v Z}eljko Ivezi{\'c}, Andrew J. Connolly, Jacob T. VanderPlas and Alexander Gray for doing the same with $\tt astroML$ \citep{astroML}. We employed plotting routines from both these packages.
%
%
%
%
\bibliographystyle{mnras}
\bibliography{example} {}

\begin{thebibliography}{}
\makeatletter
\relax
\def\mn@urlcharsother{\let\do\@makeother \do\$\do\&\do\#\do\^\do\_\do\%\do\~}
\def\mn@doi{\begingroup\mn@urlcharsother \@ifnextchar [ {\mn@doi@}
  {\mn@doi@[]}}
\def\mn@doi@[#1]#2{\def\@tempa{#1}\ifx\@tempa\@empty \href
  {http://dx.doi.org/#2} {doi:#2}\else \href {http://dx.doi.org/#2} {#1}\fi
  \endgroup}
\def\mn@eprint#1#2{\mn@eprint@#1:#2::\@nil}
\def\mn@eprint@arXiv#1{\href {http://arxiv.org/abs/#1} {{\tt arXiv:#1}}}
\def\mn@eprint@dblp#1{\href {http://dblp.uni-trier.de/rec/bibtex/#1.xml}
  {dblp:#1}}
\def\mn@eprint@#1:#2:#3:#4\@nil{\def\@tempa {#1}\def\@tempb {#2}\def\@tempc
  {#3}\ifx \@tempc \@empty \let \@tempc \@tempb \let \@tempb \@tempa \fi \ifx
  \@tempb \@empty \def\@tempb {arXiv}\fi \@ifundefined
  {mn@eprint@\@tempb}{\@tempb:\@tempc}{\expandafter \expandafter \csname
  mn@eprint@\@tempb\endcsname \expandafter{\@tempc}}}

\bibitem[\protect\citeauthoryear{{Anders} et~al.,}{{Anders}
  et~al.}{2017}]{Cor2017}
{Anders} F.,  et~al., 2017, \mn@doi [\aap] {10.1051/0004-6361/201629363}, \href
  {http://adsabs.harvard.edu/abs/2017A%26A...600A..70A} {600, A70}

\bibitem[\protect\citeauthoryear{{Arenou} et~al.,}{{Arenou}
  et~al.}{2017}]{Are2017}
{Arenou} F.,  et~al., 2017, \mn@doi [\aap] {10.1051/0004-6361/201629895}, \href
  {http://adsabs.harvard.edu/abs/2017A%26A...599A..50A} {599, A50}

\bibitem[\protect\citeauthoryear{Bailer-Jones}{Bailer-Jones}{2015}]{Bai2015}
Bailer-Jones C. A.~L.,  2015, Publications of the Astronomical Society of the
  Pacific, 127, 994

\bibitem[\protect\citeauthoryear{{Bland-Hawthorn} \&
  {Gerhard}}{{Bland-Hawthorn} \& {Gerhard}}{2016}]{Bla2016}
{Bland-Hawthorn} J.,  {Gerhard} O.,  2016, \mn@doi [\araa]
  {10.1146/annurev-astro-081915-023441}, \href
  {http://adsabs.harvard.edu/abs/2016ARA%26A..54..529B} {54, 529}

\bibitem[\protect\citeauthoryear{{Bovy}}{{Bovy}}{2015}]{galpy}
{Bovy} J.,  2015, \mn@doi [\apjs] {10.1088/0067-0049/216/2/29}, \href
  {http://adsabs.harvard.edu/abs/2015ApJS..216...29B} {216, 29}

\bibitem[\protect\citeauthoryear{{Casagrande}, {Sch{\"o}nrich}, {Asplund},
  {Cassisi}, {Ram{\'{\i}}rez}, {Mel{\'e}ndez}, {Bensby}  \&
  {Feltzing}}{{Casagrande} et~al.}{2011}]{Cas2011}
{Casagrande} L.,  {Sch{\"o}nrich} R.,  {Asplund} M.,  {Cassisi} S.,
  {Ram{\'{\i}}rez} I.,  {Mel{\'e}ndez} J.,  {Bensby} T.,   {Feltzing} S.,
  2011, \mn@doi [\aap] {10.1051/0004-6361/201016276}, \href
  {http://adsabs.harvard.edu/abs/2011A%26A...530A.138C} {530, A138}

\bibitem[\protect\citeauthoryear{{Casagrande} et~al.,}{{Casagrande}
  et~al.}{2016}]{Cas2016}
{Casagrande} L.,  et~al., 2016, \mn@doi [\mnras] {10.1093/mnras/stv2320}, \href
  {http://adsabs.harvard.edu/abs/2016MNRAS.455..987C} {455, 987}

\bibitem[\protect\citeauthoryear{{Castelli} \& {Kurucz}}{{Castelli} \&
  {Kurucz}}{2003}]{Cas2003}
{Castelli} F.,  {Kurucz} R.~L.,  2003, in {Piskunov} N.,  {Weiss} W.~W.,
  {Gray} D.~F.,  eds,  IAU Symposium Vol. 210, Modelling of Stellar
  Atmospheres. p.~A20

\bibitem[\protect\citeauthoryear{{Chaplin} \& {Miglio}}{{Chaplin} \&
  {Miglio}}{2013}]{Cha2013}
{Chaplin} W.~J.,  {Miglio} A.,  2013, \mn@doi [\araa]
  {10.1146/annurev-astro-082812-140938}, \href
  {http://adsabs.harvard.edu/abs/2013ARA%26A..51..353C} {51, 353}

\bibitem[\protect\citeauthoryear{{Cropper} \& {Katz}}{{Cropper} \&
  {Katz}}{2011}]{Cro2011}
{Cropper} M.,  {Katz} D.,  2011, in {Turon} C.,  {Meynadier} F.,   {Arenou} F.,
   eds,  EAS Publications Series Vol. 45, EAS Publications Series. pp 181--188
  (\mn@eprint {arXiv} {1010.0419}), \mn@doi{10.1051/eas/1045031}

\bibitem[\protect\citeauthoryear{Dalton et~al.,}{Dalton et~al.}{2012}]{Dal2012}
Dalton G.,  et~al., 2012, WEAVE: the next generation wide-field spectroscopy
  facility for the William Herschel Telescope: The next generation wide-field
  spectroscopy facility for the William Herschel Telescope.
p. 84460P, \mn@doi{10.1117/12.925950}

\bibitem[\protect\citeauthoryear{{Dotter}, {Chaboyer}, {Jevremovi{\'c}},
  {Kostov}, {Baron}  \& {Ferguson}}{{Dotter} et~al.}{2008}]{Dot2008}
{Dotter} A.,  {Chaboyer} B.,  {Jevremovi{\'c}} D.,  {Kostov} V.,  {Baron} E.,
  {Ferguson} J.~W.,  2008, \mn@doi [\apjs] {10.1086/589654}, \href
  {http://adsabs.harvard.edu/abs/2008ApJS..178...89D} {178, 89}

\bibitem[\protect\citeauthoryear{{Eisenstein} et~al.,}{{Eisenstein}
  et~al.}{2011}]{Eis2011}
{Eisenstein} D.~J.,  et~al., 2011, \mn@doi [\aj] {10.1088/0004-6256/142/3/72},
  \href {http://adsabs.harvard.edu/abs/2011AJ....142...72E} {142, 72}

\bibitem[\protect\citeauthoryear{Feast, Menzies, Matsunaga  \& Whitelock}{Feast
  et~al.}{2014}]{Fea2014}
Feast M.~W.,  Menzies J.~W.,  Matsunaga N.,   Whitelock P.~A.,  2014, Nature,
  509, 342

\bibitem[\protect\citeauthoryear{{Feltzing} \& {Chiba}}{{Feltzing} \&
  {Chiba}}{2013}]{Fel2013}
{Feltzing} S.,  {Chiba} M.,  2013, \mn@doi [\nar]
  {10.1016/j.newar.2013.06.001}, \href
  {http://adsabs.harvard.edu/abs/2013NewAR..57...80F} {57, 80}

\bibitem[\protect\citeauthoryear{{Gaia Collaboration} et~al.,}{{Gaia
  Collaboration} et~al.}{2016a}]{gaia2016b}
{Gaia Collaboration} et~al., 2016a, \mn@doi [\aap]
  {10.1051/0004-6361/201629272}, \href
  {http://adsabs.harvard.edu/abs/2016A%26A...595A...1G} {595, A1}

\bibitem[\protect\citeauthoryear{{Gaia Collaboration} et~al.,}{{Gaia
  Collaboration} et~al.}{2016b}]{gaia2016a}
{Gaia Collaboration} et~al., 2016b, \mn@doi [\aap]
  {10.1051/0004-6361/201629512}, \href
  {http://adsabs.harvard.edu/abs/2016A%26A...595A...2G} {595, A2}

\bibitem[\protect\citeauthoryear{{Gilmore} et~al.,}{{Gilmore}
  et~al.}{2012}]{Gil2012}
{Gilmore} G.,  et~al., 2012, The Messenger, \href
  {http://adsabs.harvard.edu/abs/2012Msngr.147...25G} {147, 25}

\bibitem[\protect\citeauthoryear{{Grand}, {Kawata}  \& {Cropper}}{{Grand}
  et~al.}{2015}]{Gra2015}
{Grand} R.~J.~J.,  {Kawata} D.,   {Cropper} M.,  2015, \mn@doi [\mnras]
  {10.1093/mnras/stv016}, \href
  {http://adsabs.harvard.edu/abs/2015MNRAS.447.4018G} {447, 4018}

\bibitem[\protect\citeauthoryear{{Grevesse} \& {Sauval}}{{Grevesse} \&
  {Sauval}}{1998}]{Gre1998}
{Grevesse} N.,  {Sauval} A.~J.,  1998, \mn@doi [\ssr]
  {10.1023/A:1005161325181}, \href
  {http://adsabs.harvard.edu/abs/1998SSRv...85..161G} {85, 161}

\bibitem[\protect\citeauthoryear{{Hayden} et~al.,}{{Hayden}
  et~al.}{2015}]{Hay2015}
{Hayden} M.~R.,  et~al., 2015, \mn@doi [\apj] {10.1088/0004-637X/808/2/132},
  \href {http://ukads.nottingham.ac.uk/abs/2015ApJ...808..132H} {808, 132}

\bibitem[\protect\citeauthoryear{{Haywood}, {Di Matteo}, {Lehnert}, {Katz}  \&
  {G{\'o}mez}}{{Haywood} et~al.}{2013}]{Hay2013}
{Haywood} M.,  {Di Matteo} P.,  {Lehnert} M.~D.,  {Katz} D.,   {G{\'o}mez} A.,
  2013, \mn@doi [\aap] {10.1051/0004-6361/201321397}, \href
  {http://adsabs.harvard.edu/abs/2013A%26A...560A.109H} {560, A109}

\bibitem[\protect\citeauthoryear{{H{\o}g} et~al.,}{{H{\o}g}
  et~al.}{2000}]{Hog2000}
{H{\o}g} E.,  et~al., 2000, \aap, \href
  {http://adsabs.harvard.edu/abs/2000A%26A...355L..27H} {355, L27}

\bibitem[\protect\citeauthoryear{{Holmberg}, {Nordstr{\"o}m}  \&
  {Andersen}}{{Holmberg} et~al.}{2009}]{Hol2009}
{Holmberg} J.,  {Nordstr{\"o}m} B.,   {Andersen} J.,  2009, \mn@doi [\aap]
  {10.1051/0004-6361/200811191}, \href
  {http://adsabs.harvard.edu/abs/2009A%26A...501..941H} {501, 941}

\bibitem[\protect\citeauthoryear{{Kawata}, {Grand}, {Gibson}, {Casagrande},
  {Hunt}  \& {Brook}}{{Kawata} et~al.}{2017}]{Kaw2017}
{Kawata} D.,  {Grand} R.~J.~J.,  {Gibson} B.~K.,  {Casagrande} L.,  {Hunt}
  J.~A.~S.,   {Brook} C.~B.,  2017, \mn@doi [\mnras] {10.1093/mnras/stw2363},
  \href {http://adsabs.harvard.edu/abs/2017MNRAS.464..702K} {464, 702}

\bibitem[\protect\citeauthoryear{{Kelly}}{{Kelly}}{2007}]{Kel2007}
{Kelly} B.~C.,  2007, \mn@doi [\apj] {10.1086/519947}, \href
  {http://adsabs.harvard.edu/abs/2007ApJ...665.1489K} {665, 1489}

\bibitem[\protect\citeauthoryear{{Kunder} et~al.,}{{Kunder}
  et~al.}{2017}]{Kun2017}
{Kunder} A.,  et~al., 2017, \mn@doi [\aj] {10.3847/1538-3881/153/2/75}, \href
  {http://adsabs.harvard.edu/abs/2017AJ....153...75K} {153, 75}

\bibitem[\protect\citeauthoryear{{Lin}, {Dotter}  \& {Asplund}}{{Lin}
  et~al.}{2017}]{Linetal17}
{Lin} J.,  {Dotter} A.,   {Asplund} M.,  2017, submitted to \mnras, \href
  {http://adsabs.harvard.edu/} {}

\bibitem[\protect\citeauthoryear{{Lindegren} et~al.,}{{Lindegren}
  et~al.}{2016}]{Lin2016}
{Lindegren} L.,  et~al., 2016, \mn@doi [\aap] {10.1051/0004-6361/201628714},
  \href {http://adsabs.harvard.edu/abs/2016A%26A...595A...4L} {595, A4}

\bibitem[\protect\citeauthoryear{{Liu} et~al.,}{{Liu} et~al.}{2014}]{Liu2014}
{Liu} C.,  et~al., 2014, \mn@doi [\apj] {10.1088/0004-637X/790/2/110}, \href
  {http://ads.bao.ac.cn/abs/2014ApJ...790..110L} {790, 110}

\bibitem[\protect\citeauthoryear{{Mackereth} et~al.,}{{Mackereth}
  et~al.}{2017}]{Mac2017}
{Mackereth} J.~T.,  et~al., 2017, preprint, \href
  {http://adsabs.harvard.edu/abs/2017arXiv170600018M} {} (\mn@eprint {arXiv}
  {1706.00018})

\bibitem[\protect\citeauthoryear{{Majewski} et~al.,}{{Majewski}
  et~al.}{2015}]{Maj2015}
{Majewski} S.~R.,  et~al., 2015, preprint, \href
  {http://adsabs.harvard.edu/abs/2015arXiv150905420M} {} (\mn@eprint {arXiv}
  {1509.05420})

\bibitem[\protect\citeauthoryear{{Martell} et~al.,}{{Martell}
  et~al.}{2017}]{Mar2017}
{Martell} S.~L.,  et~al., 2017, \mn@doi [\mnras] {10.1093/mnras/stw2835}, \href
  {http://adsabs.harvard.edu/abs/2017MNRAS.465.3203M} {465, 3203}

\bibitem[\protect\citeauthoryear{{Martig}, {Minchev}  \& {Flynn}}{{Martig}
  et~al.}{2014}]{Mar2014}
{Martig} M.,  {Minchev} I.,   {Flynn} C.,  2014, \mn@doi [\mnras]
  {10.1093/mnras/stu1003}, \href
  {http://adsabs.harvard.edu/abs/2014MNRAS.442.2474M} {442, 2474}

\bibitem[\protect\citeauthoryear{Martig et~al.,}{Martig et~al.}{2016}]{Mar2016}
Martig M.,  et~al., 2016, \mn@doi [Monthly Notices of the Royal Astronomical
  Society] {10.1093/mnras/stv2830}, 456, 3655

\bibitem[\protect\citeauthoryear{{Masseron} \& {Gilmore}}{{Masseron} \&
  {Gilmore}}{2015}]{Mas2015}
{Masseron} T.,  {Gilmore} G.,  2015, {Carbon, nitrogen and {$\alpha$}-element
  abundances determine the formation sequence of the Galactic thick and thin
  discs} (\mn@eprint {arXiv} {1503.00537}), \mn@doi{10.1093/mnras/stv1731}

\bibitem[\protect\citeauthoryear{{McMillan} et~al.,}{{McMillan}
  et~al.}{2017}]{PCM2017}
{McMillan} P.~J.,  et~al., 2017, preprint, \href
  {http://adsabs.harvard.edu/abs/2017arXiv170704554M} {} (\mn@eprint {arXiv}
  {1707.04554})

\bibitem[\protect\citeauthoryear{{Michalik}, {Lindegren}  \&
  {Hobbs}}{{Michalik} et~al.}{2015}]{Mic2015}
{Michalik} D.,  {Lindegren} L.,   {Hobbs} D.,  2015, \mn@doi [\aap]
  {10.1051/0004-6361/201425310}, \href
  {http://adsabs.harvard.edu/abs/2015A%26A...574A.115M} {574, A115}

\bibitem[\protect\citeauthoryear{{Minchev}, {Chiappini}  \& {Martig}}{{Minchev}
  et~al.}{2013}]{Min2013}
{Minchev} I.,  {Chiappini} C.,   {Martig} M.,  2013, \mn@doi [\aap]
  {10.1051/0004-6361/201220189}, \href
  {http://cdsads.u-strasbg.fr/abs/2013A%26A...558A...9M} {558, A9}

\bibitem[\protect\citeauthoryear{{Minchev}, {Chiappini}  \& {Martig}}{{Minchev}
  et~al.}{2014}]{Min2014}
{Minchev} I.,  {Chiappini} C.,   {Martig} M.,  2014, \mn@doi [\aap]
  {10.1051/0004-6361/201423487}, \href
  {http://cdsads.u-strasbg.fr/abs/2014A%26A...572A..92M} {572, A92}

\bibitem[\protect\citeauthoryear{{Minchev}, {Martig}, {Streich}, {Scannapieco},
  {de Jong}  \& {Steinmetz}}{{Minchev} et~al.}{2015}]{Min20151}
{Minchev} I.,  {Martig} M.,  {Streich} D.,  {Scannapieco} C.,  {de Jong} R.~S.,
    {Steinmetz} M.,  2015, \mn@doi [\apjl] {10.1088/2041-8205/804/1/L9}, \href
  {http://adsabs.harvard.edu/abs/2015ApJ...804L...9M} {804, L9}

\bibitem[\protect\citeauthoryear{{Minchev}, {Steinmetz}, {Chiappini}, {Martig},
  {Anders}, {Matijevic}  \& {de Jong}}{{Minchev} et~al.}{2017}]{Min2017}
{Minchev} I.,  {Steinmetz} M.,  {Chiappini} C.,  {Martig} M.,  {Anders} F.,
  {Matijevic} G.,   {de Jong} R.~S.,  2017, \mn@doi [\apj]
  {10.3847/1538-4357/834/1/27}, \href
  {http://adsabs.harvard.edu/abs/2017ApJ...834...27M} {834, 27}

\bibitem[\protect\citeauthoryear{Nakanishi \& Sofue}{Nakanishi \&
  Sofue}{2016}]{Nak2016}
Nakanishi H.,  Sofue Y.,  2016, \mn@doi [Publications of the Astronomical
  Society of Japan] {10.1093/pasj/psv108}, 68, 5

\bibitem[\protect\citeauthoryear{{Nieva} \& {Przybilla}}{{Nieva} \&
  {Przybilla}}{2012}]{Nie2012}
{Nieva} M.-F.,  {Przybilla} N.,  2012, \mn@doi [\aap]
  {10.1051/0004-6361/201118158}, \href
  {http://adsabs.harvard.edu/abs/2012A%26A...539A.143N} {539, A143}

\bibitem[\protect\citeauthoryear{{Rahimi}, {Carrell}  \& {Kawata}}{{Rahimi}
  et~al.}{2014}]{Rah2014}
{Rahimi} A.,  {Carrell} K.,   {Kawata} D.,  2014, \mn@doi [Research in
  Astronomy and Astrophysics] {10.1088/1674-4527/14/11/004}, \href
  {http://adsabs.harvard.edu/abs/2014RAA....14.1406R} {14, 1406}

\bibitem[\protect\citeauthoryear{{Ro{\v s}kar}, {Debattista}, {Quinn},
  {Stinson}  \& {Wadsley}}{{Ro{\v s}kar} et~al.}{2008}]{Ros2008}
{Ro{\v s}kar} R.,  {Debattista} V.~P.,  {Quinn} T.~R.,  {Stinson} G.~S.,
  {Wadsley} J.,  2008, \mn@doi [\apjl] {10.1086/592231}, \href
  {http://adsabs.harvard.edu/abs/2008ApJ...684L..79R} {684, L79}

\bibitem[\protect\citeauthoryear{{Ruchti} et~al.,}{{Ruchti}
  et~al.}{2015}]{Ruc2015}
{Ruchti} G.~R.,  et~al., 2015, \mn@doi [\mnras] {10.1093/mnras/stv807}, \href
  {http://adsabs.harvard.edu/abs/2015MNRAS.450.2874R} {450, 2874}

\bibitem[\protect\citeauthoryear{{Salpeter}}{{Salpeter}}{1955}]{Sal1955}
{Salpeter} E.~E.,  1955, \mn@doi [\apj] {10.1086/145971}, \href
  {http://adsabs.harvard.edu/abs/1955ApJ...121..161S} {121, 161}

\bibitem[\protect\citeauthoryear{{S{\'a}nchez-Bl{\'a}zquez}, {Courty}, {Gibson}
   \& {Brook}}{{S{\'a}nchez-Bl{\'a}zquez} et~al.}{2009}]{San2009}
{S{\'a}nchez-Bl{\'a}zquez} P.,  {Courty} S.,  {Gibson} B.~K.,   {Brook} C.~B.,
  2009, \mn@doi [\mnras] {10.1111/j.1365-2966.2009.15133.x}, \href
  {http://adsabs.harvard.edu/abs/2009MNRAS.398..591S} {398, 591}

\bibitem[\protect\citeauthoryear{{Schlesinger} et~al.,}{{Schlesinger}
  et~al.}{2014}]{Sch2014}
{Schlesinger} K.~J.,  et~al., 2014, \mn@doi [\apj]
  {10.1088/0004-637X/791/2/112}, \href
  {http://adsabs.harvard.edu/abs/2014ApJ...791..112S} {791, 112}

\bibitem[\protect\citeauthoryear{{Sch{\"o}nrich} \& {Binney}}{{Sch{\"o}nrich}
  \& {Binney}}{2009a}]{S2009b}
{Sch{\"o}nrich} R.,  {Binney} J.,  2009a, \mn@doi [\mnras]
  {10.1111/j.1365-2966.2009.14750.x}, \href
  {http://adsabs.harvard.edu/abs/2009MNRAS.396..203S} {396, 203}

\bibitem[\protect\citeauthoryear{{Sch{\"o}nrich} \& {Binney}}{{Sch{\"o}nrich}
  \& {Binney}}{2009b}]{S2009}
{Sch{\"o}nrich} R.,  {Binney} J.,  2009b, \mn@doi [\mnras]
  {10.1111/j.1365-2966.2009.15365.x}, \href
  {http://adsabs.harvard.edu/abs/2009MNRAS.399.1145S} {399, 1145}

\bibitem[\protect\citeauthoryear{{Sch{\"o}nrich} \& {McMillan}}{{Sch{\"o}nrich}
  \& {McMillan}}{2017}]{Sch2017}
{Sch{\"o}nrich} R.,  {McMillan} P.~J.,  2017, \mn@doi [\mnras]
  {10.1093/mnras/stx093}, \href
  {http://adsabs.harvard.edu/abs/2017MNRAS.467.1154S} {467, 1154}

\bibitem[\protect\citeauthoryear{{Scoville}, {Thakkar}, {Carlstrom}  \&
  {Sargent}}{{Scoville} et~al.}{1993}]{Sco1993}
{Scoville} N.~Z.,  {Thakkar} D.,  {Carlstrom} J.~E.,   {Sargent} A.~I.,  1993,
  \mn@doi [\apjl] {10.1086/186743}, \href
  {http://adsabs.harvard.edu/abs/1993ApJ...404L..59S} {404, L59}

\bibitem[\protect\citeauthoryear{{Sellwood} \& {Binney}}{{Sellwood} \&
  {Binney}}{2002}]{Sel2000}
{Sellwood} J.~A.,  {Binney} J.~J.,  2002, \mn@doi [\mnras]
  {10.1046/j.1365-8711.2002.05806.x}, \href
  {http://adsabs.harvard.edu/abs/2002MNRAS.336..785S} {336, 785}

\bibitem[\protect\citeauthoryear{{Smecker-Hane} \& {Wyse}}{{Smecker-Hane} \&
  {Wyse}}{1992}]{Sme1992}
{Smecker-Hane} T.~A.,  {Wyse} R.~F.~G.,  1992, \mn@doi [\aj] {10.1086/116175},
  \href {http://adsabs.harvard.edu/abs/1992AJ....103.1621S} {103, 1621}

\bibitem[\protect\citeauthoryear{{Steinmetz} et~al.,}{{Steinmetz}
  et~al.}{2006}]{Ste2006}
{Steinmetz} M.,  et~al., 2006, \mn@doi [\aj] {10.1086/506564}, \href
  {http://adsabs.harvard.edu/abs/2006AJ....132.1645S} {132, 1645}

\bibitem[\protect\citeauthoryear{{Stello} et~al.,}{{Stello}
  et~al.}{2015}]{Ste2015}
{Stello} D.,  et~al., 2015, \mn@doi [\apjl] {10.1088/2041-8205/809/1/L3}, \href
  {http://adsabs.harvard.edu/abs/2015ApJ...809L...3S} {809, L3}

\bibitem[\protect\citeauthoryear{{Vanderplas}, {Connolly}, {Ivezi{\'c}}  \&
  {Gray}}{{Vanderplas} et~al.}{2012}]{astroML}
{Vanderplas} J.,  {Connolly} A.,  {Ivezi{\'c}} {\v Z}.,   {Gray} A.,  2012, in
  Conference on Intelligent Data Understanding (CIDU). pp 47 --54,
  \mn@doi{10.1109/CIDU.2012.6382200}

\bibitem[\protect\citeauthoryear{{Wojno} et~al.,}{{Wojno}
  et~al.}{2016}]{Woj2016}
{Wojno} J.,  et~al., 2016, preprint, \href
  {http://adsabs.harvard.edu/abs/2016arXiv161100733W} {} (\mn@eprint {arXiv}
  {1611.00733})

\bibitem[\protect\citeauthoryear{{Xiang} et~al.,}{{Xiang}
  et~al.}{2015}]{Xia2015}
{Xiang} M.-S.,  et~al., 2015, \mn@doi [Research in Astronomy and Astrophysics]
  {10.1088/1674-4527/15/8/009}, \href
  {http://adsabs.harvard.edu/abs/2015RAA....15.1209X} {15, 1209}

\bibitem[\protect\citeauthoryear{{Yanny} et~al.,}{{Yanny}
  et~al.}{2009}]{Yanny2009}
{Yanny} B.,  et~al., 2009, \mn@doi [\aj] {10.1088/0004-6256/137/5/4377}, \href
  {http://adsabs.harvard.edu/abs/2009AJ....137.4377Y} {137, 4377}

\bibitem[\protect\citeauthoryear{{de Jong} et~al.,}{{de Jong}
  et~al.}{2012}]{deJong2012}
{de Jong} R.~S.,  et~al., 2012, in Ground-based and Airborne Instrumentation
  for Astronomy IV. p. 84460T (\mn@eprint {arXiv} {1206.6885}),
  \mn@doi{10.1117/12.926239}

\makeatother
\end{thebibliography}
%
%
%
%
%
\clearpage
\appendix
\section{Comparison analysis with PJM2017}
\label{appendix}

\begin{figure}
	\centering
	\includegraphics[width=\columnwidth]{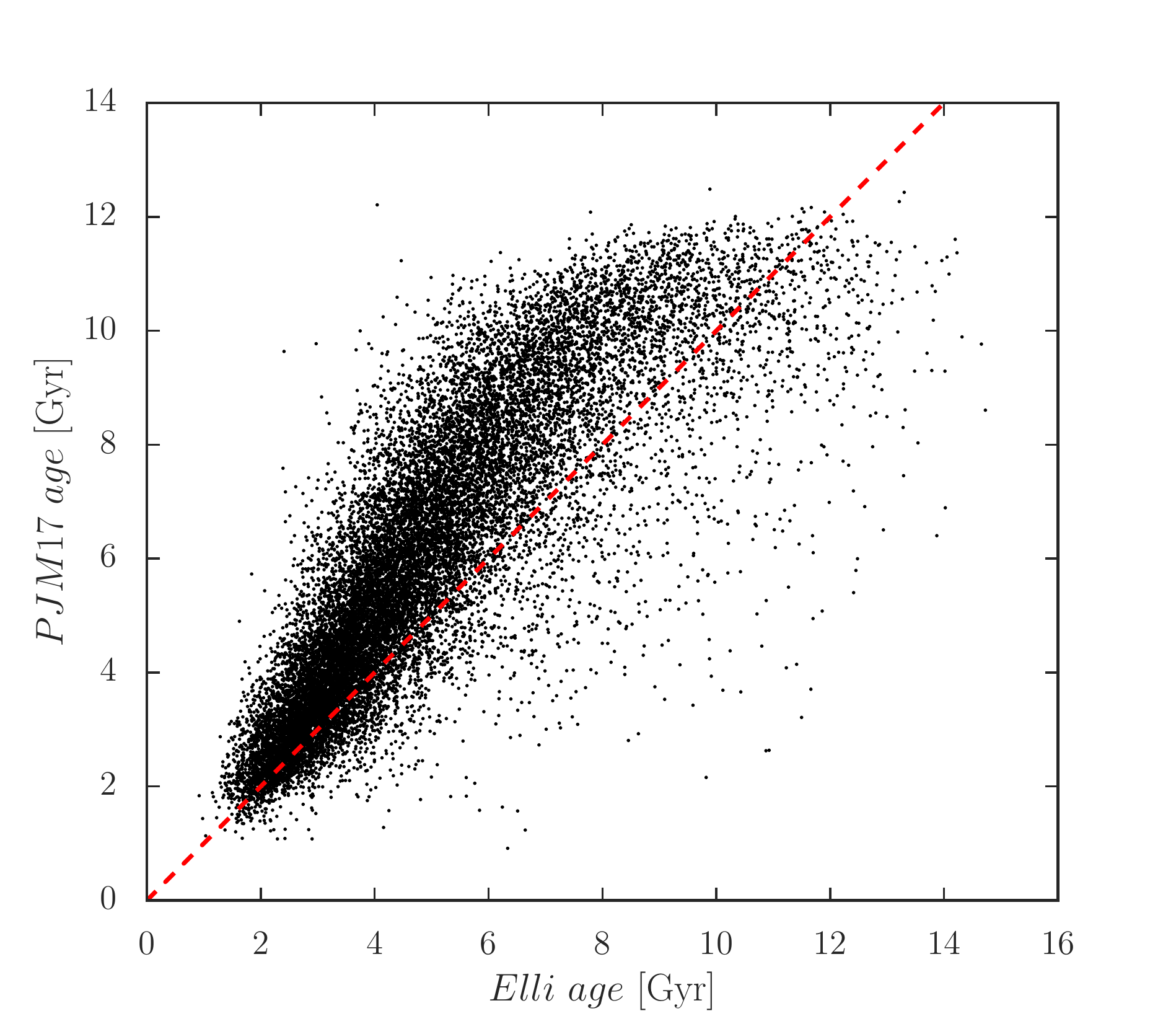}
	\caption{Comparison between ages derived using $\tt Elli$ and the  $\tt PJM17$ method for the sample of stars used in this paper (see Sec. \ref{method}). The red dotted line displays a 1:1 relationship and it can be seen that the $\tt PJM17$ age estimates are systematically larger than those obtained using $\tt Elli$. }
	\label{comparison}
\end{figure}

\begin{figure}
	\centering
	\includegraphics[width=\columnwidth]{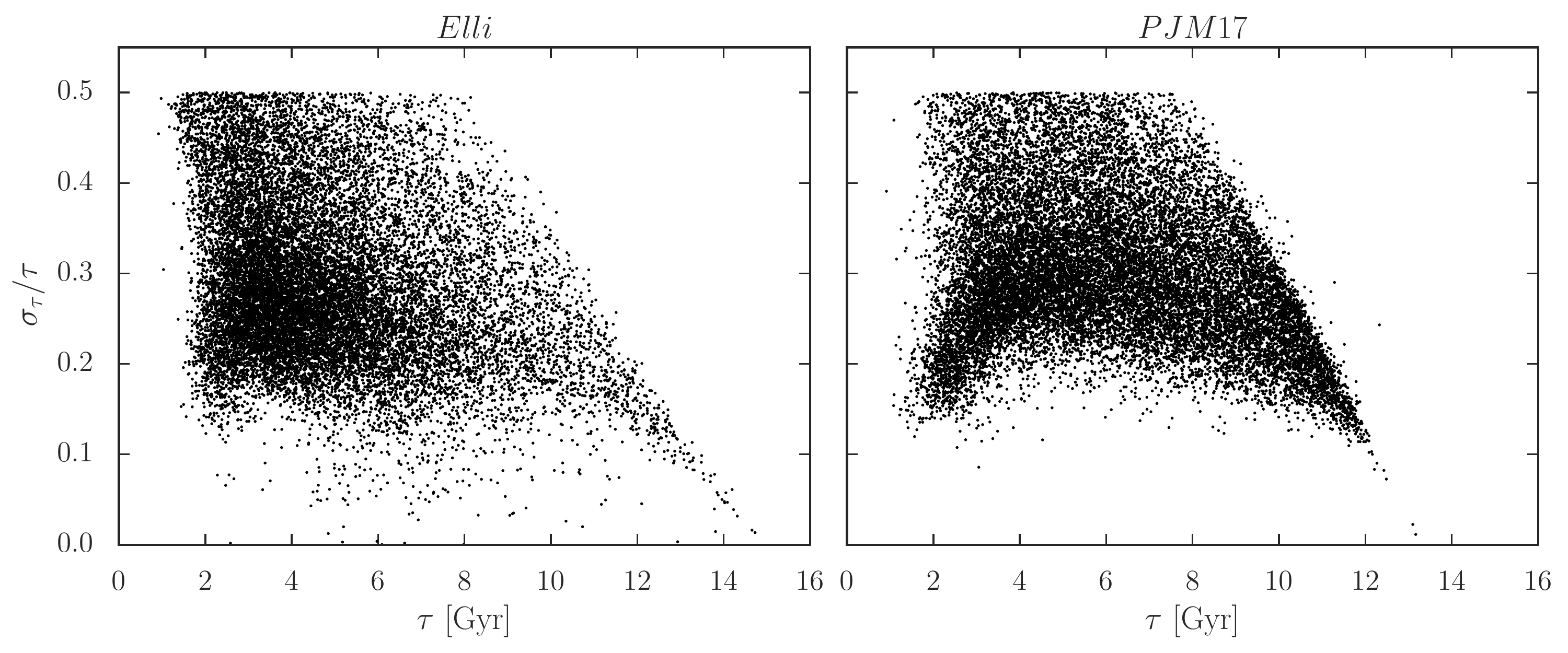}
	\caption{Fractional age uncertainty as a function of age for both the ``Elli sample" and ``PJM2017 sample" (see text).}
	\label{frac_err}
\end{figure}

In this paper, we used the age estimate obtained using the isochrone fitting method $\tt Elli$ \citep{Linetal17}. To test if our results are not affected by choice of the technique employed to get age estimates, we have applied the same analysis to the age estimates provided in \cite{PCM2017}, hereafter $\tt PJM17$, who used a different isochrone fitting method and different isochrones for the TGAS-RAVE sample. First, we compare the ages derived with $\tt Elli$ and $\tt PJM17$ in Fig. \ref{comparison}. The $\tt PJM17$ method provides overall larger derived ages than $\tt Elli$ and the difference between these two age estimates is more significant in the older population.

\begin{figure*}
	\centering
	\includegraphics[width=\textwidth]{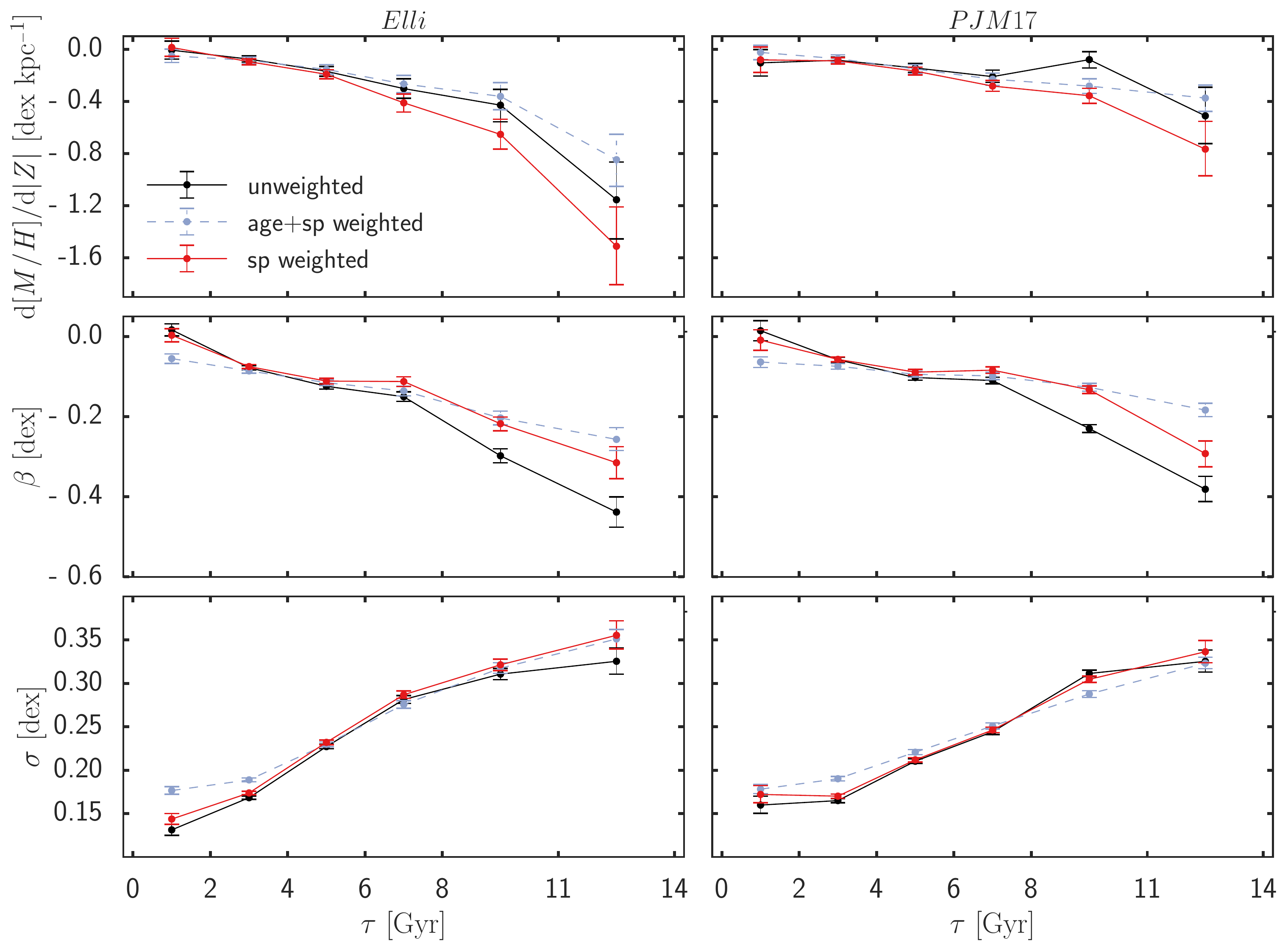}
	\caption{Comparison between the results for slope, intercept and dispersion of the vertical metallicity profile of stars  from the $\tt Elli$ (left panel) and $\tt PJM17$ (right panel) samples (see text). The unweighted estimates are shown in black, the stellar population weighted ones in red, and the age-weighted ones in light blue.}
	\label{final_comp}
\end{figure*}

We applied the same analysis described in Sec. \ref{method} to the dataset using the $\tt PJM17$ age estimates and their uncertainties. Furthermore, we also cut in relative age error at less than 50 \%. A subsequent MCMC approach is taken to characterize the slope, intercept and dispersion of the vertical metallicity profile of the mono-age populations considered in the main text. For this comparison, we also discuss the results obtained for stars older than 11~Gyr because $\tt PJM17$ provides a statistically significant sample of these stars. The last age bin contains stars older than 11~Gyr in Fig. \ref{final_comp}, but for visualisation, we plot the results at age=14~Gyr. We also applied the distance estimated in $\tt PJM17$ rather than the distance from the TGAS parallax to show that the results are not affected by choice of distance estimate. We call this sample the ``PJM2017 sample" and the sample selected in Sec. \ref{method} the ``$Elli$ sample". Fractional age uncertainty as a function of age for both samples is shown in Fig. \ref{frac_err}. Both samples are comparable in terms of the distribution of age uncertainties against age.

Fig. \ref{final_comp} shows the results of the stellar population bias weighted (noted as ``sp-weighted") and unweighted (noted as ``unweighted") cases. Within their estimated uncertainties, the results for the gradient of the metallicity as a function of height for the unweighted case are consistent between these analyses, except the age bins of 0-2~Gyr and 8-11~Gyr where the $\tt PJM17$ sample shows rather unexpectedly steeper and flat gradient, respectively. The results from the $\tt PJM17$ sample preserve the main conclusion of this paper qualitatively, namely that older stars have a more negative vertical metallicity gradient than younger ones, with the stellar population weighted result showing it more clearly.

Interestingly, the results for the slope for the 8<$\tau$<11~Gyr age bin found from our analysis using $\tt Elli$ are similar to those obtained for the >11~Gyr age bin in the $\tt PJM17$ sample. This trend is present across all ages considered, with the results for the vertical metallicity gradient obtained for the $\tt Elli$ sample being comparable to those for the $\tt PJM17$ sample in one age bin older. In other words, we can argue that although the age estimate may suffer from the systematic uncertainty as demonstrated in Fig. \ref{comparison}, we can trust the relative trend of vertical metallicity gradient with respect to stellar age. Hence, this comparison shows that the qualitative trend we obtained from the analysis using the $\tt Elli$ sample is not affected by the systematic biases or errors in the isochrone age estimates. 

Fig. \ref{final_comp} also provides the results taking into account the age uncertainties by computing the weights of the contribution to each age bin for each star following \cite{Cor2017} and assuming the age is drawn from a Gaussian distribution. We also weigh the age weights by the stellar population described in Section \ref{:method}. The results are labeled as ``age-sp weighted" in Fig. \ref{final_comp}. We applied the same weighting procedure for the logarithm of the likelihood function as we did for the stellar population bias weights in Sec. \ref{method}. Fig. \ref{final_comp} shows that the slopes in the age-sp-weighted case are systematically flatter as stars with different mean ages with large errors contribute to adjacent age bins. Therefore, we think that the age-weighted results are underestimating the slope.  Nevertheless, there is still a qualitative trend of more negative slopes in older mono-age populations. This also provides the robustness of our conclusions. Fig. \ref{final_comp} also presents the results of the intercept and the dispersion. The trends are generally consistent with the results of the slope, and again this demonstrates the robustness of our conclusion.

%
%
%
\bsp	
\label{lastpage}
\end{document}